\documentclass[11pt]{article}

\usepackage{amsmath}
\usepackage{graphicx}
\usepackage{amsfonts}
\usepackage{amssymb}
\usepackage{epsfig}
\usepackage{color}
\usepackage{psfrag}
\usepackage{epstopdf}

\newcommand{\EQ}{\begin{equation}}
\newcommand{\EN}{\end{equation}}
\newcommand{\be}{\begin{equation}}
\newcommand{\ee}{\end{equation}}
\newcommand{\bea}{\begin{eqnarray}}
\newcommand{\eea}{\end{eqnarray}}

\begin{document} \setcounter{page}{0}
\topmargin 0pt
\oddsidemargin 5mm
\renewcommand{\thefootnote}{\arabic{footnote}}
\newpage
\setcounter{page}{0}
\topmargin 0pt
\oddsidemargin 5mm
\renewcommand{\thefootnote}{\arabic{footnote}}
\newpage
\begin{titlepage}
\begin{flushright}
\end{flushright}
\vspace{0.5cm}
\begin{center}
{\large {\bf Particles, conformal invariance and criticality\\
 in pure and disordered systems}}\\
\vspace{1.8cm}
{\large Gesualdo Delfino}\\
\vspace{0.5cm}
{\em SISSA -- Via Bonomea 265, 34136 Trieste, Italy}\\
{\em INFN sezione di Trieste}\\
\end{center}
\vspace{1.2cm}

\renewcommand{\thefootnote}{\arabic{footnote}}
\setcounter{footnote}{0}

\begin{abstract}
\noindent
The two-dimensional case occupies a special position in the theory of critical phenomena due to the exact results provided by lattice solutions and, directly in the continuum, by the infinite-dimensional character of the conformal algebra. However, some sectors of the theory, and most notably criticality in systems with quenched disorder and short range interactions, have appeared out of reach of exact methods and lacked the insight coming from analytical solutions. In this article we review recent progress achieved implementing conformal invariance within the particle description of field theory. The formalism yields exact unitarity equations whose solutions classify critical points with a given symmetry. It provides new insight in the case of pure systems, as well as the first exact access to criticality in presence of short range quenched disorder. Analytical mechanisms emerge that in the random case allow the superuniversality of some critical exponents and make explicit the softening of first order transitions by disorder.
\end{abstract}
\end{titlepage}

\newpage

\tableofcontents

\section{Introduction}
\label{introduction}
The fact that statistical systems can admit points of their parameter space at which the correlation length diverges allows to cast within a same theoretical framework problems as diverse as magnetism, phase transitions in fluids, percolation, polymer satistics, or superconductivity (see e.g. \cite{Cardy_book}). These {\it critical points} correspond to fixed points of the renormalization trasformations that inspect how the system responds to a change of scale \cite{WK}. While the need to deal with infinitely many degrees of freedom makes of critical phenomena an intrinsically difficult problem that, in general, can only be approached within suitable approximation schemes, a special circumstance turns the two-dimensional case into a forward position. Field theory, which is the framework one is led to in the study of critical phenomena, shows that scale invariance at critical points actually results into invariance under the larger group of {\it conformal} transformations \cite{Polyakov,DfMS}. In dimension $d=2$ the conformal group has infinitely many generators, and this has led to an impressive body of exact results starting from the representation theory of the conformal algebra \cite{DfMS,BPZ}. 

On the other hand, these powerful methods, which for conciseness we can call "algebraic", could not yet provide all the insight about critical phenomena that one would like to extract exploiting the special character of the two-dimensional case. On the side of structural properties, a pending case has been that of geometrical criticality, of which percolation provides a main example \cite{SA}. In contrast with the problem of boundary connectivity \cite{Cardy92,Smirnov}, the probability that $k$ points in the bulk belong to a same percolation cluster \cite{DV_4point} is not determined by the differential equations provided by the algebraic framework. Only in recent years the way to answer the question for $k=3$ has been found \cite{DV_3point}, stimulating the study of the case $k=4$ \cite{PRS,JS_perc,Dotsenko_perc,HJS,NR}, which is sensitive to the whole spectrum of scaling dimensions of the theory\footnote{See \cite{DV_4point,GV} for the difference with the boundary $k=4$ case.}. 

The questions that we consider in this article arise at an earlier stage, since they are concerned with the very existence of critical points, and with the problem of gaining theoretical access to them. A first illustration is obtained recalling that criticality is usually associated to spontaneous symmetry breaking, a circumstance that makes clear the role of internal symmetry in the classification of critical phenomena. This role, however, is not always easy to combine with the complex algebraic structure associated to spatial (i.e. conformal) symmetry. Hence, for example, the question concerning the maximal value $q_\textrm{max}$ for which criticality can be achieved in Potts models (permutational symmetry $\mathbb{S}_q$) has traditionally remained unanswered even in two dimensions. Lattice results \cite{Baxter} are sufficient to determine a maximal value $4$ for ferromagnets, for which the lattice structure is immaterial, but leave room for larger values to be realized in antiferromagnets, which require a case by case study.

Even more general has been the problem with random criticality. This concerns systems with {\it quenched disorder}, i.e. including degrees of freedom (e.g. impurities) that take too long to reach thermal equilibrium and on observable time scales behave as random variables (see \cite{Cardy_book}). Experiments and numerical simulations show that these systems admit critical points with critical exponents differing from those of the "pure" systems (i.e. without disorder). In two dimensions, it has also been generally known that disorder is able to soften first order phase transitions of the pure models into second order ones \cite{AW,HB}, thus making more room for criticality and conformal invariance. This, however, only made more strident the total absence of exact results for random criticality\footnote{Throughout the paper we always consider the case of short range interactions.}. Actually, given the inability of establishing an analytical link, the relevance of conformal invariance for random criticality could only be argued on numerical grounds \cite{AHHM,BLdM,JLdPSW,KW}. 

Making progress on such longstanding problems requires a formalism that, while taking advantage of the infinite-dimensional character of conformal symmetry in $d=2$, remains able to provide exact informations in presence of generic symmetries or quenched disorder. It was shown in \cite{paraf,random} that this is possible turning to the {\it particle} framework. The fact that field theory also allows for a particle description is very well known (see e.g. \cite{Ryder} for an introduction), but has not been traditionally used to deal with critical points. The reason is that the particle formalism relies on the analytic properties of the probability amplitudes for scattering processes \cite{ELOP}. The particles, which in statistical field theory correspond to collective excitation modes, have a mass $m$ inversely proportional to the correlation length $\xi$. At criticality $\xi=\infty$, and the vanishing of the mass causes infinitely many branch points of a scattering amplitude to collapse onto each other, making the analytic structure intractable. This infrared catastrophe, however, does not occur in $d=2$, since conformal symmetry yields infinitely many conservation laws for the scattering and leaves only a finite number of branch cuts. Hence, the analytic properties remain under control. In addition, scale invariance forces the amplitudes to be energy-independent and allows their exact determination \cite{paraf}, which amounts to a classification of critical points with a given internal symmetry. 

For pure systems, this classification includes the known critical points, but can give access to new ones. Hence, remaining on the example of the $q$-state Potts model, one finds that in absence of disorder $q_\textrm{max}=(7+\sqrt{17})/2$ \cite{DT1}, which leaves room for criticality in a $q=5$ antiferromagnet. The value of $q_\textrm{max}$ illustrates that the framework accounts for analytic continuation to noninteger values of $q$, a possibility known from the cluster expansion on the lattice \cite{FK}. This ability to perform analytic continuations in the symmetry parameters is also a key for accessing random criticality, since it crucially enters the replica representation of quenched disorder \cite{Cardy_book}. The particle framework then allows to explore the space of random fixed points and, for example, to see analytically for the first time that disorder moves $q_\textrm{max}$ to infinity  \cite{random}. Exact access to random criticality also reveals unsuspected theoretical mechanisms. Indeed, some solutions of the fixed point equations turn out to exhibit sectors that do not depend on symmetry parameters, and then allow for some {\it superuniversal} critical exponents. In particular, this explains \cite{random} why, for the most studied critical line of the random bond Potts model, numerical and experimental investigations \cite{CFL,DW,KSSD,random_exp,CJ,CB2,OY,JP,Jacobsen_multiscaling,AdAI} always found $\nu\approx 1$ for the correlation length exponent in the range $q\in[2,\infty)$, while clear $q$-dependence was observed for other exponents in \cite{CJ} and later studies. The results provided by the particle framework  finally show that conformal invariance, although exploited in a way very different from the traditional one, is able to give new and essential insight also about random criticality.

The presentation is organized as follows. In the next section we recall some general notions of critical phenomena and the related vocabulary. In section~3 we turn to $d=2$ and recall some results of conformal field theory, as well as the properties of the line of Gaussian fixed points. In section~4 we introduce particles and the scattering formalism, and show how to use them at critical points. Section~5 illustrates the particle formalism for the case of pure systems, through the analysis of two main models of the theory of critical phenomena, the $O(N)$ vector model and the $q$-state Potts model. The same is then done for the random case in section~6, after recalling some general aspects of quenched disorder and its theoretical treatment within the replica setting. Some final remarks are given in section~7.

\section{Notions of critical phenomena}
\label{general}
We consider systems of equilibrium statistical mechanics \cite{LL} specified by their Hamiltonian ${\cal H}$, for which the expectation value of an observable ${\cal O}$ is the statistical average over configurations
\EQ
\langle{\cal O}\rangle=\frac{1}{Z}\sum_\textrm{configurations}{\cal O}\,e^{-{\cal H}/T},
\label{average}
\EN
where $T$ is the temperature and
\EQ
Z=\sum_\textrm{configurations}e^{-{\cal H}/T}
\label{pf}
\EN
is the partition function. It is convenient, to start with, to look at the degrees of freedom of a system as variables $s_i$ ("spins") located at sites $i$ of a regular lattice that provides a discretization of the $d$-dimensional Euclidean space $\mathbb{R}^d$, with the sums in (\ref{average}) and (\ref{pf}) corresponding to sums over all spin configurations. We will refer to systems with short range interactions among the site variables. Interactions are also homogeneous throughout the lattice, so that, in particular, $\langle s_i\rangle$ is site-independent. 

In general, the site variable $s_i$ has several components and carries a representation of the symmetry group $G$ that leaves invariant the Hamiltonian\footnote{Since we will focus on critical systems, we directly consider Hamiltonians invariant under the symmetry group $G$ of the critical point.}. A basic example is the vector model with Hamiltonian
\EQ
{\cal H}_\textrm{vector}=-J\sum_{\langle i,j\rangle}{\bf s}_i\cdot{\bf s}_j\,,
\label{vector}
\EN
where ${\bf s}_i$ is a $N$-component unit vector, so that $G=O(N)$, and the sum is taken over all nearest neighbor pairs of sites $\langle i,j\rangle$. The case $N=1$ ($G=\mathbb{Z}_2$) defines the Ising model. The coupling $J$ distinguishes between ferromagnetic ($J>0$) and antiferromagnetic ($J<0$) interactions; for the time being we will refer to ferromagnets. A different generalization of the Ising model is provided by the $q$-state Potts model  \cite{Potts,Wu} defined by the Hamiltonian 
\EQ
{\cal H}_\textrm{Potts}=-J\sum_{\langle i,j\rangle}\delta_{s_i,s_j}\,,
\label{potts}
\EN
where $s_i=1,2,\ldots,q$. The Hamiltonian is invariant under global permutations of the $q$ values of the site variables (often referred to as "colors"), so that the symmetry corresponds to the permutational group $S_q$. The $\mathbb{Z}_2$ symmetry characteristic of the Ising model is recovered for $q=2$. 
 
Going back to the general discussion, the "order parameter" $\langle s_i\rangle$ (or a suitable variant\footnote{In the Potts model one takes $\langle\sigma_{\alpha,i}\rangle=\langle\delta_{s_i,\alpha}-\frac{1}{q}\rangle$, $\alpha=1,2,\ldots,q$.}) vanishes for $T$ large enough, and becomes nonzero when the temperature is lowered below a critical value $T_c$, which is a phase transition point corresponding to the spontaneous breaking of the symmetry $G$. Below $T_c$ the order parameter can take different values related by the symmetry. The transition is said to be of the first order if the order parameter is discontinuous at $T_c$, and  of the second order (or, more generally, continuous) otherwise. 

The decay
\EQ
\langle s_is_j\rangle-\langle s_i\rangle^2\sim e^{-|i-j|/\xi}
\label{xi}
\EN
of the spin-spin correlation function for large separation $|i-j|$ between the two sites defines the correlation length $\xi$, which provides the characteristic scale of the system. The correlation length remains finite at a first order transition, but diverges as 
\EQ
\xi\sim |T-T_c|^{-\nu}\,,\hspace{1cm}T\to T_c
\label{xi_scaling}
\EN
when approaching a second order transition point; this relation defines the correlation length critical exponent $\nu$. The divergence of $\xi$ leaves no characteristic scale at distances much larger than lattice spacing, and leads to {\it scale invariance} at $T_c$. Unless otherwise specified, when talking about critical points and critical properties we will refer to second order transition points and to the properties resulting from scale invariance at these points. Thermodynamical observables exhibit scaling close to criticality, and for the order parameter this takes the form 
\EQ
\langle s_i\rangle\sim (T_c-T)^\beta\,,\hspace{1cm}T\to T_c^-\,,
\label{magnetization}
\EN
which defines the critical exponent $\beta$. 

The fact that $\xi\to\infty$ as $T\to T_c$ allows a continuum description of the near-critical region, as long as one is interested in the properties of the system over scales much larger than lattice spacing (see e.g. \cite{Cardy_book}). Such a continuum description corresponds to a field theory. The lattice variable $s_i$ is replaced by a spin field $s(x)$, where $x=(x_1,\ldots,x_d)$ denotes a point in $\mathbb{R}^d$. Similarly, the local spin-spin interaction $\sum_j s_i s_j$, with the sum taken over neighbors of the site $i$, corresponds to an energy density field $\varepsilon(x)$. The sums in (\ref{average}) and (\ref{pf}) are now taken over field configurations. In the continuum, the Hamiltonian ${\cal H}$ is invariant under the group $G$ of internal symmetry, but also under continuous spatial translations (homogeneity) and rotations (isotropy). 

At criticality, scale invariance leads to power law decay of correlation functions, and for a field $\Phi(x)$ one has the expression
\EQ
\langle\Phi(x_1)\Phi(x_2)\rangle=\frac{\textrm{constant}}{|x_1-x_2|^{2X_\Phi}}\,,
\label{power}
\EN
which defines the {\it scaling dimension} $X_\Phi$ of the field; it implies that, from the dimensional point of view, the field behaves as $\textrm{length}^{-X_\Phi}$. 

We will use the notation ${\cal A}={\cal H}/T$ for the reduced Hamiltonian, or Euclidean action, which characterizes a given theory. From the theoretical point of view, it is useful to separate the near-critical action ${\cal A}$ into a scale invariant part ${\cal A}^*$ corresponding to $T_c$, plus a term accounting for the deviation from $T_c$, namely
\EQ
{\cal A}={\cal A}^*+\tau\int d^d x\,\varepsilon(x)\,,
\label{scaling}
\EN
where $\tau\sim T-T_c$. Since ${\cal A}$ is dimensionless (recall (\ref{average})), $\tau$ has the dimension of an inverse length to the power $d-X_\varepsilon$ and provides the dimensionful coupling that breaks scale invariance away from criticality. Since $\xi$ is a length, we have $\xi\propto|\tau|^{-1/(d-X_\varepsilon)}$ and, comparing with (\ref{xi_scaling}), the critical exponent 
\EQ
\nu=1/(d-X_\varepsilon)\,.
\label{nu}
\EN

If $\Phi(x)$ is a field with given transformation properties under the action of the symmetry group $G$, the field theory contains infinitely many fields with growing scaling dimension and transforming in the same way (it is sufficient to think to the derivatives of $\Phi(x)$). In writing (\ref{scaling}) we omitted the contribution of infinitely many $G$-invariant fields with scaling dimensions larger than $d$. The couplings conjugated to such fields have the dimension of a length to positive powers and become negligible for the description of the large distance properties of our interest. It is in this sense that such fields are called ``irrelevant'' in the language of the renormalization group (RG) that expresses the response of the system to a change of scale \cite{Cardy_book,WK}. A scale invariant theory is called a RG {\it fixed point}. The action in which irrelevant fields are omitted and which describes the large distance properties is called the scaling action. It may contain more than one $G$-invariant field with scaling dimension smaller than $d$ (these fields are called ``relevant''). This occurs when more than one parameter needs to be tuned to achieve scale invariance; such theories describe ``multicritical'' behavior. In light of these considerations, the field that we denote $\varepsilon(x)$ can be more precisely defined as the most relevant (smallest scaling dimension) $G$-invariant field; similarly, the spin field $s(x)$ is the most relevant field with the symmetry properties of the order parameter. Some theories also possess fields with scaling dimension equal to $d$ (``marginal'' fields). Marginality may be spoiled by logarithmic corrections induced by interaction. Depending on the theory, these corrections will effectively produce a "marginally relevant" or a "marginally irrelevant" field. If no logarithmic correction occurs, the addition to a fixed point action of such a ``truly marginal'' field does not break scale  invariance and generates a line of fixed points.

Notice also that the order parameter scales as $\langle s(x)\rangle\sim\xi^{-X_s}\sim|\tau|^{\nu X_s}$, so that comparison with (\ref{magnetization}) yields
\EQ
\beta=\nu X_s=X_s/(d-X_\varepsilon)\,.
\label{beta}
\EN
The relations (\ref{nu}) and (\ref{beta}) illustrate the fact that the critical exponents are determined by the scaling dimensions, which then are the fundamental critical indices.

\section{Aspects of critical behavior in two dimensions}
\label{conformal}
\subsection{Some results from conformal symmetry}
\label{minimal}
Let us now turn to critical systems in $d=2$. In writing the correlation function (\ref{power}) we considered the simplest case in which the field $\Phi$ is scalar, namely is invariant under spatial rotations. More generally, we can consider fields $\Phi(x)$ with scaling dimension $X_\Phi$ that transform as 
\EQ
\Phi(0)\to e^{-is_\Phi\alpha}\Phi(0)
\label{rotation}
\EN
under a rotation by an angle $\alpha$ centered in the origin; $s_\Phi$ is called the "Euclidean spin" of the field. A very important property of field theory is that products of fields can be expanded onto an infinite-dimensional basis of fields \cite{WK}. In a scale invariant theory, the requirement that the result preserves the dimensional and rotational properties implies for such an operator product expansion (OPE) the form
\bea
\Phi_i(x)\Phi_j(0)&=&\sum_k C_{ij}^{k}\,(z\bar{z})^{(X_k-X_i-X_j)/2}(z\bar{z}^{-1})^{(s_k-s_i-s_j)/2}\,\Phi_k(0)\nonumber\\
&=&\sum_k C_{ij}^{k}\,z^{\Delta_{k}-\Delta_{i}-\Delta_{j}}\bar{z}^{\bar{\Delta}_{k}-\bar{\Delta}_{i}-\bar{\Delta}_{j}}\,\Phi_k(0)\,,
\label{ope}
\eea
where we introduced the complex coordinates on the plane
\EQ
z=x_1+ix_2\,,\hspace{.9cm}\bar{z} = x_1-ix_2\,,
\EN
which transform under rotations as  $z\to e^{i\alpha}z$, $\bar{z}\to e^{-i\alpha}\bar{z}$, as well as $\Delta_\Phi$, $\bar{\Delta}_\Phi$ such that 
\bea
&& X_\Phi=\Delta_\Phi+\bar{\Delta}_\Phi\,,
\label{X_phi}\\
&& s_\Phi=\Delta_\Phi-\bar{\Delta}_\Phi\,.
\label{s_phi}
\eea
In (\ref{ope}) the $C_{ij}^k$'s are called OPE coefficients, and we used the shortened notations $X_{\Phi_i}=X_i$,~...~. Notice that the final expression of (\ref{ope}) allows to treat a field $\Phi(x)$ as the product of a $z$-dependent part with dimension $\Delta_\Phi$, and a $\bar{z}$-dependent part with dimension $\bar{\Delta}_\Phi$. The OPE also allows to introduce the useful notion of mutual locality. In ordinary physical cases one expects the correlation functions $\langle\cdots\Phi_i(x)\Phi_j(0)\cdots\rangle$ to be invariant if $x$ is taken around the origin and brought to the original position, namely under the continuation $z\to e^{2i\pi}z$, $\bar{z}\to e^{-2i\pi}\bar{z}$. If this is the case, the fields $\Phi_i$ and $\Phi_j$ are said to be mutually local. The OPE (\ref{ope}) shows that the condition is satisfied if
\EQ
s_i+s_j-s_k\in \mathbb{Z}\,
\label{gamma}
\EN
for all $k$'s in the sum.

An important advantage of the field theoretical framework is that it allows to show (see e.g. \cite{DfMS}) that scale invariant theories are actually invariant under the larger group of conformal transformations, i.e. tranformations in which the change of scale, instead of being global, varies smoothly with the coordinate. Due to this property, the scale invariant field theories describing statistical systems at criticality actually correspond to {\it conformal field theories} (CFTs). This circumstance has its most powerful implications in the two-dimensional case of interest in this article, since in this case conformal transformations correspond to variations $\delta z=f(z)$, $\delta\bar{z}=\bar{f}(\bar{z})$, where $f$ (resp. $\bar{f}$) is any analytic function of $z$ (resp. $\bar{z}$). It follows that in $d=2$ the conformal group has the essential peculiarity of possessing infinitely many generators. These turn out to correspond \cite{DfMS,BPZ} to operators $L_n$ satisfying the Virasoro algebra
\EQ
[L_n,L_m]=(n-m)L_{n+m}+\frac{c}{12}(n^3-n)\delta_{n,-m}\,,
\label{Virasoro}
\EN
where $c$ is a key parameter of the critical system known as "central charge". We now recall some of the implications of the algebra (\ref{Virasoro}), referring the reader to \cite{DfMS,BPZ} for the derivations. In the first place, the space of fields in critical theories splits into families corresponding to lowest weight representations of the algebra. Considering the $z$-dependent part (a similar structure holds for the $\bar{z}$-dependent part), a family $[\phi]$ contains a "primary" field $\phi(z)$ with dimension (also known as conformal dimension) $\Delta_\phi$, together with "descendants" with dimension $\Delta_\phi+l$; $l=1,2,\ldots$ is the "level" of the descendant. The derivatives of $\phi$ are examples of descendants. An important role is played by the reducible representations of the algebra, i.e. representations $[\phi]$ that contain another representation $[\phi_0]$ whose primary $\phi_0$ is a descendant of $\phi$ at a level $l_0$. The irreducible representation that one obtains factoring out $[\phi_0]$ is said to be "degenerate" at level $l_0$, and $\phi$ is said to be a degenerate primary. The construction can be shown to lead to differential equations for multi-point correlation functions containing a degenerate primary.

The most advanced exploration of the space of CFTs is available for central charge $c\leq 1$. For $c<1$ it is convenient to use the parameterization  
\EQ
c=1-\frac{6}{p(p+1)}\,,
\label{c_p}
\EN
with $p>0$. Then the degenerate primaries can be written as $\Phi_{m,n}(z)$, with $m$ and $n$ positive integers, and their dimension $\Delta_{m,n}$ is determined by\footnote{For later convenience, in (\ref{deltamunu}) we use indices $\mu,\nu$ that are not necessarily integer.}
\EQ
\Delta_{\mu,\nu}=\frac{[(p+1)\mu-p\nu]^2-1}{4p(p+1)}\,.
\label{deltamunu}
\EN
The OPE of two degenerate primaries takes the form
\EQ
\Phi_{m_1,n_1}\cdot\Phi_{m_2,n_2}=\sum_{k=0}^{min(m_1,m_2)-1}\,\sum_{l=0}^{min(n_1,n_2)-1}\left[\Phi_{|m_1-m_2|+1+2k,|n_1-n_2|+1+2l}\right]\,,
\label{dd}
\EN
where we have suppressed the coordinate dependence, which is generally determined by (\ref{ope}); the notation $[\Phi]$ in the r.h.s. indicates contribution from the whole family of fields. As we saw, a field is the product of a $z$-dependent part and a $\bar{z}$-dependent part, and (\ref{dd}) separately applies to each of them. One can also consider fields $\Phi_{\mu,\nu}(z)$ with dimension (\ref{deltamunu}) and noninteger indices; they are nondegenerate and satisfy the OPE
\EQ
\Phi_{m,n}\cdot\Phi_{\mu,\nu}=\sum_{k=0}^{m-1}\,\sum_{l=0}^{n-1}\left[\Phi_{\mu-m+1+2k,\nu-n+1+2l}\right]\,
\label{dnd}
\EN
with degenerate fields.  

Ordinary critical points satisfy "reflection positivity", which implies a spectrum of conformal dimension $\{\Delta_i\}$ without negative values, so that correlations decay with distance. Remarkably, for $c<1$ reflection positivity is satisfied only by (\ref{c_p}) with $p=3,4,\ldots$ \cite{FQS}. For these values of $c$ the OPE (\ref{dd}) closes on a {\it finite} number of families originating from degenerate primaries \cite{BPZ}, giving rise to the so called reflection positive "minimal models". For these theories the conformal dimensions $\Delta_{m,n}$ of the primary fields are given by (\ref{deltamunu}) with $m=1,2,\ldots,p-1$, $n=1,2,\ldots,p$. Notice that the number of primaries, and then the number of families, grows with $c$. It is a general property of reflection positive CFTs that the central charge gives a measure of the number of degrees of freedom \cite{Z_cth}. 

As expected, the minimal field content ($p=3$, $c=1/2$) corresponds to the Ising critical point, with $\Delta_{1,1}=\Delta_{2,3}=0$, $\Delta_{1,2}=\Delta_{2,2}=1/16$ and $\Delta_{1,3}=\Delta_{2,1}=1/2$ corresponding to the identity, the spin field and the energy density field, respectively. These scalar fields have $\Delta=\bar{\Delta}$, and the scaling dimensions $X_s=1/8$ and $X_\varepsilon=1$ determine through (\ref{nu}) and (\ref{beta}) the Ising critical exponents, in agreement with lattice results \cite{Onsager,Yang}. The Ising spectrum of conformal dimensions also allows to build the fermions (i.e. fields with half-integer spin) $\psi$ and $\bar{\psi}$ with dimensions $(\Delta,\bar{\Delta})$ equal to $(1/2,0)$ and $(0,1/2)$, respectively. Since dimension 0 corresponds to the identity, and does not carry a coordinate dependence, we have\footnote{We use the notation $\partial=\partial_z$ and $\bar{\partial}=\bar{\partial}_{\bar{z}}$.} $\bar{\partial}\psi=\partial\bar{\psi}=0$, which are the equations of motion of free fermions. It is known since the lattice results of \cite{Kaufman} that the two-dimensional Ising model (without external field) corresponds to a free fermionic theory.

The minimal models with $p>3$ generally correspond to multicritical points associated to the spontaneous breaking of $\mathbb{Z}_2$ symmetry \cite{Zamo_multicritical,Huse}; in particular, $p=4$ yields the tricritical Ising model\footnote{Tricriticality can be realized, for example, in a generalized Ising model allowing for vacant sites.}. However, the values $p=5,6$ also allow a restriction to a smaller set of primaries \cite{Dotsenko,Cardy_modular,CIZ} which yields the critical and tricritical three-state Potts ferromagnet, respectively.

\subsection{Gaussian critical line}
\label{c1}
Gaussian criticality in two dimensions corresponds to the theory of a free scalar field with action
\EQ
{\cal A}_0=\frac{1}{4\pi}\int d^2x\,(\nabla\varphi)^2\,.
\label{free}
\EN
While in $d>2$ the Gaussian fixed point is synonym of triviality, the two-dimensional case is much more interesting. The fact that $\varphi(x)$ is dimensionless leads to the logarithmic correlator
\EQ
\langle\varphi(x)\varphi(0)\rangle=-\ln|x|=-\frac{1}{2}(\ln z+\ln\bar{z})\,,
\label{log}
\EN
which is consistent with the equation of motion $\partial\bar{\partial}\varphi=0$ and the decomposition 
\EQ
\varphi(x)=\phi(z)+\bar{\phi}(\bar{z})\,.
\EN
Instead of $\phi$, which has $\langle\phi(z)\phi(0)\rangle\propto\ln z$, proper scaling primary fields of the theory are the exponentials
\EQ
V_{p}(z)=e^{2ip\phi(z)}\,,
\label{V}
\EN
 whose dimension easily follows from free field methods (see \cite{DfMS}) and reads
\EQ
\Delta_{V_p}\equiv\Delta_p=p^2\,.
\label{deltap}
\EN
Of course one also has $\bar{V}_{\bar{p}}(\bar{z})=e^{2i\bar{p}\bar{\phi}(\bar{z})}$, in such a way that the generic primary $V_p\bar{V}_{\bar{p}}$ has dimensions $(\Delta_p,\Delta_{\bar{p}})$. The Gaussian OPE has the form
\EQ
V_{p_1}\cdot V_{p_2}=\left[V_{p_1+p_2}\right]\,,
\label{gaussian}
\EN
and, together with (\ref{gamma}) and (\ref{deltap}), implies that two fields $V_{p_1}\bar{V}_{\bar{p}_1}$ and $V_{p_2}\bar{V}_{\bar{p}_2}$ are mutually local if 
\EQ
2(p_1p_2-\bar{p}_1\bar{p}_2)\in\mathbb{Z}\,.
\label{gaussian_locality}
\EN
A generic choice for the energy density field in this theory is $\varepsilon=V_b\bar{V}_b+V_{-b}\bar{V}_{-b}\propto\cos 2b\varphi$, with $\Delta_\varepsilon=\bar{\Delta}_\varepsilon=b^2$, and physically interesting fields are local with respect to $\varepsilon$. A field $V_p\bar{V}_{\bar{p}}$ satisfies this condition if $\pm 2b(p-\bar{p})$ is an integer, i.e. if 
\EQ
p-\bar{p}=\frac{m}{2b}\,,\hspace{1cm}m\in\mathbb{Z}\,.
\label{quantization}
\EN
For $m=1$ we can build the complex fermion 
\EQ
\Psi=(\psi,\bar{\psi})=\left(V_{\frac{1}{4b}+\frac{b}{2}}\bar{V}_{-\frac{1}{4b}+\frac{b}{2}},V_{\frac{1}{4b}-\frac{b}{2}}\bar{V}_{-\frac{1}{4b}-\frac{b}{2}}\right)\,,
\label{dirac}
\EN
with spin $p^2-\bar{p}^2$ equal to $1/2$ for $\psi$ and to $-1/2$ for $\bar{\psi}$. The decomposition $\Psi=\Psi_1+i\Psi_2$ defines two real fermions $\Psi_i=(\psi_i,\bar{\psi}_i)$. When $b^2=1/2$ we have $\bar{\partial}\psi=\partial{\bar \psi}=0$, which are free fermionic equations of motion; it follows that for $b^2=1/2$ the theory (\ref{free}) can be represented in terms of free fermions. On the other hand, for $b^2\neq 1/2$ the fermions are coupled by the four-fermion term, which can be shown to be truly marginal; it follows that the action (\ref{free}) with energy density field $\cos 2b\varphi$ can be expressed as \cite{Coleman,Mandelstam}
\EQ
{\cal A}_0=\int d^2x\,\left[\sum_{i=1,2}(\psi_i\bar{\partial}\psi_i+\bar{\psi}_i\partial\bar{\psi_i})+g(b^2)\psi_1\bar{\psi}_1\psi_2\bar{\psi_2}\right]\,,
\label{Thirring}
\EN
with $g(1/2)=0$; the field $\cos2b\varphi$ corresponds to the fermionic mass term $\psi_1\bar{\psi_1}+\psi_2\bar{\psi_2}$. The form (\ref{Thirring}) of the action makes transparent that the two-dimensional Gaussian model actually corresponds to a {\it line a fixed points} parameterized by $b^2$. Since at $b^2=1/2$ we have two free neutral fermions, namely two decoupled Ising models, the central charge is twice the Ising one, namely $c=1$. This value, on the other hand, holds generically for the Gaussian model, since the interaction ($g\neq 0$) does not change the field content. The theory (\ref{Thirring}) quite directly describes the critical properties of the Ashkin-Teller model \cite{AT,KB}, corresponding to two Ising models coupled by energy-energy interaction. This model indeed possesses a critical line along which critical exponents vary continuously.

The existence of a scalar (bosonic) description (\ref{free}) and a fermionic description (\ref{Thirring}) for a same theory is a remarkable property of the two-dimensional case. It also allows to unveil a symmetry that is not obvious in the representation (\ref{free}). Indeed, the fermionic property $\psi_i^2=\bar{\psi}_i^2=0$ allows to write $\psi_1\bar{\psi_1}\psi_2\bar{\psi_2}\propto(\sum_i\psi_i\bar{\psi}_i)^2$. As a consequence, the action (\ref{Thirring}) is left invariant by $O(2)$ rotations of the vector $(\psi_1,\psi_2)$; these are $U(1)$ transformations for the complex fermion, and the integer $m$ in (\ref{quantization}) corresponds to the $U(1)$ charge. We see that, while for central charge $c<1$ we only found discrete internal symmetries, the case $c=1$ allows for the simplest continuous symmetry, $G=O(2)\sim U(1)$. The Gaussian model then describes the critical properties of the $N=2$ vector model (\ref{vector}), which is also known as $XY$ model. The components of the spin field ${\bf s}(x)=(s_1(x),s_2(x))$ are obtained picking up the scalar fields with $m=\pm 1$, namely
\EQ
s_\pm=s_1\pm is_2=V_{\pm 1/4b}\bar{V}_{\mp 1/4b}\,,
\label{XYspin}
\EN
with $\Delta_{s_\pm}=1/16b^2$. While continuous symmetries do not break spontaneously in two dimensions \cite{MW,Hohenberg,Coleman_goldstone}, the $XY$ model exhibits a different type of transition known as Berezinskii-Kosterlitz-Thouless (BKT) transition \cite{Berezinskii,KT}. The order parameter $\langle{\bf s}(x)\rangle$ vanishes at all temperatures, but a value $T_{BKT}$ separates a high temperature phase with exponential decay of correlations from a low temperature phase (BKT phase) with power law decay. The Gaussian model naturally accounts for this phenomenon upon identification of $b^2$ as a decreasing function of the temperature (see e.g. \cite{Cardy_book}). The transition is driven by the $O(2)$-invariant ($m=0$) field $\varepsilon$, which has $X_\varepsilon=2b^2$: for $b^2>b^2(T_{BKT})=1$ the field is irrelevant and scale invariance is preserved at large distances, thus explaining the BKT phase. Since we know from (\ref{XYspin}) that $\Delta_s=1/16$ at $b^2=1$, we have 
\EQ
\langle {\bf s}(x)\cdot{\bf s}(0)\rangle_{T=T_{BKT}}\sim |x|^{-1/4}\,. 
\label{XYdecay}
\EN
For later use we are also interested in the chiral (i.e. with $\Delta=0$ or $\bar{\Delta}=0$) fields that satisfy (\ref{quantization}) and have lowest charge $m=\pm 1$; they correspond to $\eta_\pm=V_{\pm 1/2b}$ and $\bar{\eta}_\pm=\bar{V}_{\pm 1/2b}$, with 
\EQ
\Delta_{\eta_\pm}=\bar{\Delta}_{\bar{\eta}_\pm}=\frac{1}{4b^2}\,
\label{XYchiral} 
\EN
as nonzero conformal dimensions.

\section{Particle description}
\label{particles}
\subsection{Notions of scattering theory}
A field theory describing the near-critical properties of a statistical system also possesses a formulation in momentum space. Within such a formulation, the fundamendal degrees of freedom are the particle modes describing the excitations with respect to the ground state. It is worth stressing that these particle modes describe {\it collective excitations} of the degrees of freedom in real space, and should not be identified with individual particles (atoms, molecules) of a fluid system. As already noted, we are discussing field theories possessing translation and rotation invariance (Euclidean field theories), and in $d=2$ we denote by $x=(x_1,x_2)$ a point in real space. Upon the identification 
\EQ
x_2=it\,,
\label{Wick}
\EN
an Euclidean field theory defines a quantum field theory with spatial coordinate $x_1$ and time coordinate $t$. The quantum theory has the same field content as the Euclidean theory, and correlation functions in the two cases are related by the analytic continuation (\ref{Wick}). The particles modes correspond to excitations above a minimum energy (vacuum) state $|0\rangle$. Spontaneous symmetry breaking of the internal symmetry $G$ manifests through to the fact that the vacuum is not unique below $T_c$ (see e.g. \cite{Ryder}). Upon the continuation (\ref{Wick}), rotation invariance (isotropy) in Euclidean space is mapped onto relativistic invariance in space-time. Hence, the particle modes have the relativistic dispersion relation 
\EQ
E=\sqrt{p^2+m^2}\,,
\label{relativistic}
\EN 
where $E$, $p$ and $m$ are the particle energy, momentum and mass, respectively. In our natural units, the mass has the dimension of an inverse length, and is related to the correlation length as
\EQ
\xi\propto 1/m\,.
\label{xi_m}
\EN

\begin{figure}
\begin{center}
\includegraphics[width=3cm]{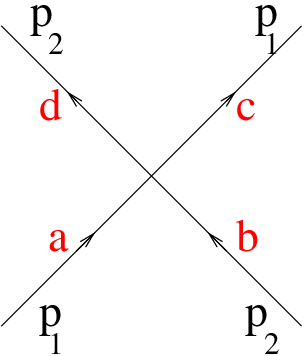}
\caption{Scattering process corresponding to the amplitude $S_{ab}^{cd}(s)$. Time runs upwards.}
\label{fpu_elastic}
\end{center} 
\end{figure}

The particle description of field theory is encoded in the $S$-matrix \cite{Ryder,ELOP}, whose matrix elements are the probability amplitudes that a set of particles at $t=-\infty$ evolves into a set of particles at $t=+\infty$ upon scattering. Relativistic scattering conserves in general total energy and momentum, but not the number of particles. Given an initial state, the sum of the transition probabilities over all possible final states has to be one, implying the {\it unitarity} of the $S$-matrix. A scattering process in which the number of particles is preserved is called "elastic". A two-particle elastic process is depicted in figure~\ref{fpu_elastic}; we consider particles with the same mass, and in $d=2$ conservation of energy and momentum implies that the momenta $p_1$ and $p_2$ are individually conserved. In general, the particles form multiplets carrying a representation of the group $G$ of internal symmetry, and the indices $a,b,c,d$ in the figure label components of the multiplets.

\begin{figure}
\begin{center}
\includegraphics[width=12cm]{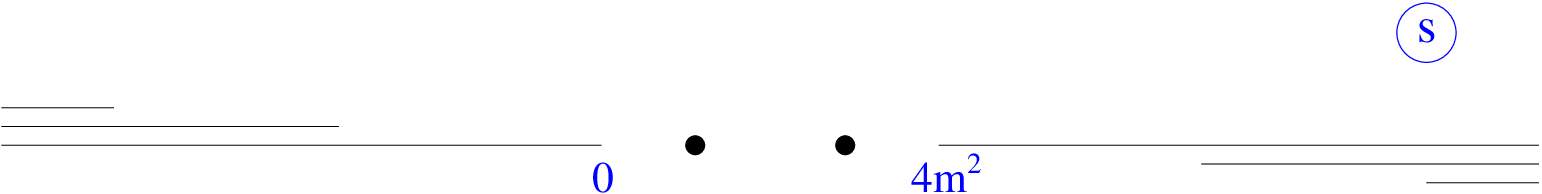}
\caption{Analytic structure of the scattering amplitude in the complex $s$-plane. The unitarity (right) and crossing (left) branch cuts are shown together with two poles.}
\label{fpu_s_plane}
\end{center} 
\end{figure}

The scattering amplitude of figure~\ref{fpu_elastic} depends on the single relativistic invariant that can be built out of the two energy-momenta, namely the square of the center of mass energy
\EQ
s=(E_1+E_2)^2-(p_1+p_2)^2\,.
\label{e2}
\EN
The amplitude, that we denote by $S_{ab}^{cd}(s)$, satisfies the relations
\bea
&& S_{ab}^{cd}(s)=S_{ba}^{dc}(s)\,,\\
\label{parity}
&& S_{ab}^{cd}(s)=S_{cd}^{ab}(s)\,,
\label{timereversal}
\eea
which express invariance under spatial inversion and time reversal, respectively. The amplitude also satisfies 
analytic properties \cite{ELOP} that we now recall. Upon continuation to complex values of the variable $s$, the amplitude is an analytic function whose singularities have a physical meaning. As a consequence of unitarity, the minimal energy values (thresholds) $s=(km)^2$ needed to produce a final state with $k\geq 2$ particles correspond to branch points of the amplitude. Instead, a pole at $s=\tilde{m}^2\in(0,4m^2)$ corresponds to a particle of mass $\tilde{m}$ appearing as a bound state of the two scattering particles. The unitarity branch cuts are taken along the positive real axis in the complex $s$-plane (figure~\ref{fpu_s_plane}), and the physical values of the amplitude are given by the limit towards the real axis from above (${S}_{ab}^{cd}(s+i\epsilon)$ with $s\geq 4m^2$, $\epsilon\to 0^+$). These values lie on the first (called ``physical'') sheet of the cut $s$-plane, other sheets being accessed through the branch cuts. The unitarity condition reads
\EQ
\sum_{e,f}{S}_{ab}^{ef}(s+i\epsilon)\left[{S}_{ef}^{cd}(s+i\epsilon)\right]^*=\delta_{ac}\delta_{bd}\,,\hspace{1cm}(2m)^2<s<s_1\,,
\label{unitarity0}
\EN
when $s$ does not exceed the first inelastic threshold $s_1$.

A further property of relativistic scattering is {\it crossing symmetry}, stating that the amplitude for the direct scattering channel (figure~\ref{fpu_elastic} with time running upwards) is related by analytic continuation to the amplitude for the crossed scattering channel (figure~\ref{fpu_elastic} with time running from left to right). When passing to the crossed channel, the particles $b$ and $d$, whose arrows point in the 'wrong' direction, are replaced with antiparticles $\bar{b}$ and $\bar{d}$, and their energy and momentum is reversed ($E_2,p_2\to-E_2,-p_2$, corresponding to $s\to 4m^2-s$). As a consequence, the crossing relation reads
\EQ
{S}_{ab}^{cd}(s+i\epsilon) = {S}_{\bar{d}a}^{\bar{b}c}(4m^2-s-i\epsilon)\,,
\label{crossing0}
\EN
with $s$ real. This relation implies that an amplitude acquires crossed channel branch cuts running along the negative real axis, together with crossing images of bound state poles (figure~\ref{fpu_s_plane}). 

Finally, the property of "real analyticity" states that the values of the amplitude on opposite edges of a cut are related by complex conjugation, 
\EQ
{S}_{ab}^{cd}(s+i\epsilon) =\left[{S}_{ab}^{cd}(s-i\epsilon)\right]^*\,.
\label{ra}
\EN

When a field $\phi$ acts on the vacuum state $|0\rangle$, it creates a state that can be expanded on the basis of multi-particle states. If this expansion includes the one-particle state $|p\rangle$, i.e. if 
\EQ
F_\phi(p)\equiv\langle p|\phi(0)|0\rangle\neq 0\,,
\label{creation}
\EN
one says that $\phi$ creates the particle; the usual normalization condition $\langle p_1|p_2\rangle=2\pi E_1\delta(p_1-p_2)$ holds. Since rotations in Euclidean space correspond to relativistic transformations in space-time under which $(E,p)$ transforms as a vector, $E\pm p$ has Euclidean spin $\pm 1$. It then follows from (\ref{rotation}), (\ref{X_phi}) and (\ref{s_phi}) that
\EQ
F_\phi(p)=a_\phi (E+p)^{\Delta_\phi}(E-p)^{\bar{\Delta}_\phi}\,,
\label{Fphi}
\EN
where $a_\phi$ is a dimensionless constant.

\subsection{Scattering at criticality}
\label{scale}
\begin{figure}
\begin{center}
\includegraphics[width=7cm]{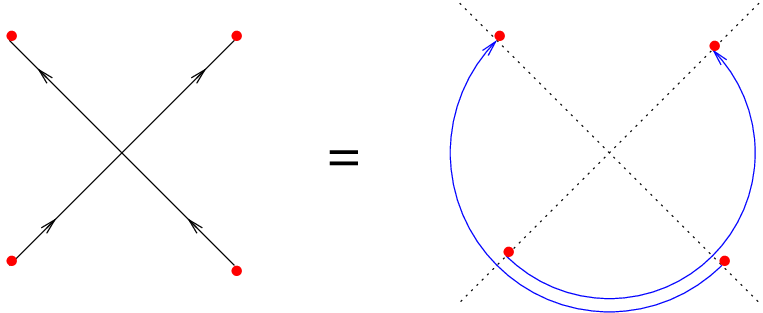}
\caption{Illustration of equation (\ref{phase}) in the $(1+1)$-dimensional space-time.}
\label{paraf_phases}
\end{center} 
\end{figure}

At a scale invariant point in two dimensions the scattering problem undergoes remarkable simplifications \cite{paraf}. In the first place, infinite-dimensional conformal symmetry implies that infinitely many quantities, and not just energy and momentum, have to be conserved in the scattering. This forces the initial and final states to be kinematically identical (same number of particles, same energies, same momenta), a property that we call "complete elasticity". In addition, since $\xi=\infty$, the particles are massless and (\ref{relativistic}) shows that their energy and momentum are related as $p=E>0$ (right movers) or $p=-E<0$ (left movers). It then follows from (\ref{creation}) and (\ref{Fphi}) that at criticality the particles are created by chiral fields $\eta(z)$ (for right movers) and $\bar{\eta}(\bar{z})$ (for left movers), i.e. fields with $\bar{\Delta}_\eta=0$, and $\Delta_{\bar{\eta}}=0$. One also has 
\EQ
s_\eta=-s_{\bar{\eta}}=\Delta_{{\eta}}=\bar{\Delta}_{\bar{\eta}}
\label{chiral}
\EN
for the spin of the fields. Since scale invariance implies the absence of dimensionful parameters, the scattering amplitude $S$ of a right-mover with a left-mover cannot depend on the variable (\ref{e2}), which is the only relativistic invariant in the process and is dimensionful. The energy-independence of the amplitude means that the particles have no dynamical interaction. This, however, does not imply $S=1$. Indeed, scattering in one spatial dimension involves position exchange on the line, and in general produces a statistical factor. This can be determined observing that, in absence of dynamical interaction, the scattering, i.e. the passage from the initial to the final state, can also be realized by $\pi$-rotations (see figure~\ref{paraf_phases}), and is then ruled by the Euclidean spin (\ref{chiral}) of the fields that create the particles. Recalling (\ref{rotation}), we obtain the scattering (statistical) amplitude \cite{paraf}
\EQ
{S}=e^{-i\pi(s_\eta-s_{\bar{\eta}})}=e^{-2i\pi\Delta_\eta}\,.
\label{phase}
\EN
We see that ${S}$ is 1 for bosons ($\Delta_\eta$ integer) and $-1$ for fermions ($\Delta_\eta$ half-integer), but can take other values corresponding to generalized statistics. 

We deduced (\ref{phase}) referring to the simplest case of a single particle species. More generally, the particles carry indices and we have the amplitudes $S_{ab}^{cd}$ of the previous subsection, with the difference that they no longer depend on the center of mass energy. With respect to the general analyticity structure of figure~\ref{fpu_s_plane}, complete elasticity eliminates all branch points, apart from the elastic one at $s=4m^2$ and its crossing image at $s=0$. Hence, the limit $m\to 0$ relevant for the present case does not involve any collapse of infinitely many branch points on top of each other, and remains well defined. Since there are no inelastic thresholds ($s_1=\infty$) and $m=0$, the unitarity equation (\ref{unitarity0}) holds for any $s$, consistently with the $s$-independence of the amplitudes; it now becomes \cite{paraf}
\EQ
\sum_{e,f}{S}_{ab}^{ef}\left[{S}_{ef}^{cd}\right]^*=\delta_{ac}\delta_{bd}\,.
\label{massless_unitarity}
\EN
On the other hand, (\ref{crossing0}) and (\ref{ra}) can be combined to obtain
\EQ
{S}_{ab}^{cd}=\left[{S}_{\bar{d}a}^{\bar{b}c}\right]^*\,.
\label{massless_crossing}
\EN
If there is a single particle species, (\ref{massless_unitarity}) yields an amplitude $S$ that is a phase, consistently with (\ref{phase}). Otherwise, a phase satisfying (\ref{phase}) can be obtained upon diagonalization of the scattering, as we are going to see in specific applications.

\section{Criticality in pure systems}
\subsection{$O(N)$ vector model}
\label{o(n)}
\subsubsection{Fixed point equations}
\begin{figure}
\begin{center}
\includegraphics[width=9cm]{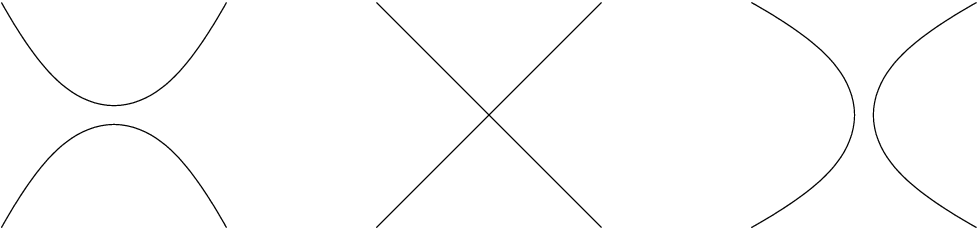}
\caption{Amplitudes ${S}_1$, ${S}_2$ and ${S}_3$ of the $O(N)$-invariant theory. Time runs upwards.}
\label{ON_ampl}
\end{center} 
\end{figure}

\begin{figure}
\begin{center}
\includegraphics[width=9cm]{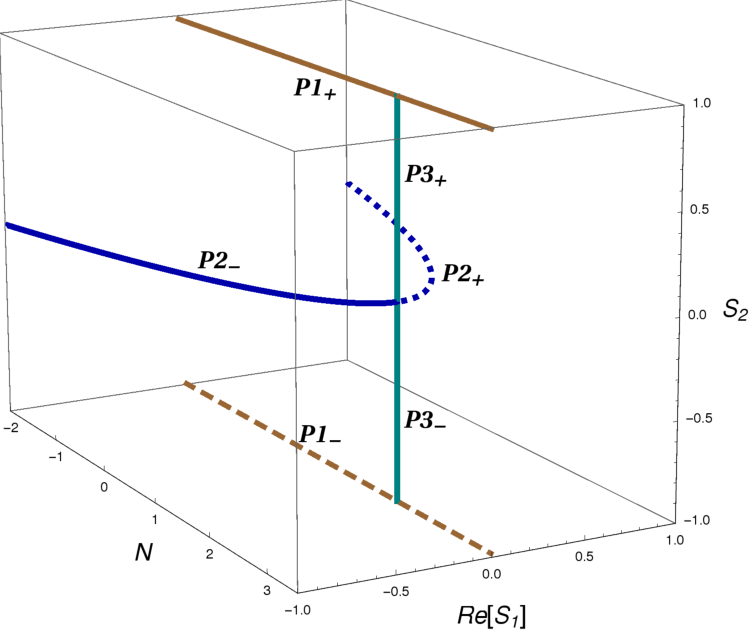}
\caption{Solutions of the fixed point equations (\ref{ONuni1})-(\ref{ONuni3}) for the $O(N)$ model. $P2_-$ and $P2_+$ correspond to the critical lines for the dilute and dense regimes of nonintersecting loops. $P3_+$ accounts for the BKT phase of the $XY$ model. $P1_+$ yields the zero temperature critical point that the model exhibits for $N>2$.
}
\label{pure_space}
\end{center} 
\end{figure}

We consider the vector model, defined on the lattice by the Hamiltonian (\ref{vector}), for a first illustration of the use of the scale invariant scattering theory of section \ref{scale}. The $O(N)$ symmetry is represented by a vector multiplet of massless particles that we denote by an index $a=1,2,\ldots,N$. We denote by $|ab\rangle$ a state containing two particles $a$ and $b$; we omit the indication of momenta since we have seen that scattering at criticality does not depend on them. Being the product of two vector representations, the initial state $|ab\rangle$ has a tensorial structure that has to be preserved by the scattering. The effect of the latter can then be written as
\EQ
|ab\rangle\to\delta_{ab}\,{S}_1\sum_{c=1}^N|cc\rangle+{S}_2\,|ab\rangle+{S}_3\,|ba\rangle\,,
\label{ON_algebra}
\EN
where $S_1$, $S_2$ and $S_3$ are annihilation, transmission and reflection amplitudes, respectively, and are depicted in figure~\ref{ON_ampl}. Since in the present case the particles are self-conjugated ($a=\bar{a}$), the crossing symmetry relations (\ref{massless_crossing}) take the form
\bea
{S}_1={S}_3^* & \equiv & \rho_1e^{i\phi}\,,\\
{S}_2={S}_2^* & \equiv &\rho_2\,,
\eea
where we have introduced parameterizations in terms of $\rho_1\geq 0$, and $\rho_2$ and $\phi$ real. In this way, the unitarity equations (\ref{massless_unitarity}) read \cite{paraf}
\bea
&& \rho_1^2+\rho_2^2=1\,,
\label{ONuni1}\\
&& \rho_1\rho_2\cos\phi=0\,,
\label{ONuni2}\\
&& N\rho_1^2+2\rho_1\rho_2\cos\phi+2\rho_1^2\cos 2\phi=0\,,
\label{ONuni3}
\eea
and correspond to the choices $(c=a, d=b)$, $(c=b, d=a)$, and $(a=b, c=d)$, respectively. Notice that in these equations $N$ enters as a parameter that does not need to be an integer. The possibility to continue the model to noninteger values of $N$ is already known on the lattice, as we recall in the next subsection. The solutions of the equations (\ref{ONuni1})-(\ref{ONuni3}) yield the critical points (RG fixed points) of the $O(N)$ model. We list them in table~\ref{ON_pure_solutions} (see also figure~\ref{pure_space}) and proceed to their discussion.

\subsubsection{Critical lines of nonintersecting loops}
\label{nil}
We begin with the solutions $P2_\pm$, which are defined for $N\in[-2,2]$. For $N=2$ they coincide with the point $S_2=0$ of the solution $P3$ that, as we will discuss in a moment, corresponds to a CFT with central charge $c=1$. Since the central charge increases with the number of degrees of freedom, and then with $N$, the CFTs corresponding to the solutions $P2_\pm$ have $c\leq 1$. We are then in the CFT subspace discussed in section~\ref{minimal}, where we have seen that a main physical role is played by the degenerate primary fields with conformal dimensions (\ref{deltamunu}). In particular, the energy density field $\varepsilon(x)$ is expected to be a degenerate field, and at $N=1$ for one of the two solutions $P2_\pm$ it should have the dimension $\Delta_\varepsilon=1/2$ of the Ising model. This leads to the identification\footnote{The alternative choice $\Delta_{2,1}$ corresponds to the $q$-state Potts model, see \cite{paraf} and below.} $\Delta_\varepsilon=\bar{\Delta}_\varepsilon=\Delta_{1,3}$. 

\begin{table}
\begin{center}
\begin{tabular}{c|c|c|c}
\hline 
Solution & $N$ & $\rho_2$ & $\cos\phi$ \\ 
\hline \hline
$\text{P}1_{\pm}$ & $(-\infty,\infty)$ & $\pm 1$ & - \\ 
$\text{P}2_{\pm}$ & $[-2, 2]$ & $0$ & $\pm\frac{1}{2}\sqrt{2-N}$ \\ 
$\text{P}3_{\pm}$ & $2$ & $\pm \sqrt{1- \rho_1^2}$ & $0$ \\[0.7em] 
\hline 
\end{tabular} 
\caption{Solutions of the Eqs.~(\ref{ONuni1})-(\ref{ONuni3}). Thy correspond to RG fixed points with $O(N)$ symmetry.
}
\label{ON_pure_solutions}
\end{center}
\end{table}

With this result, we can use the OPE (\ref{dd}) to idenfity the chiral field $\eta$ that creates the particles as the most relevant chiral field local with respect to $\varepsilon$. This leads to $\Delta_\eta=\Delta_{2,1}$, and then to the determination of $\Delta_\eta$ as a function of the parameter $p$ entering (\ref{c_p}). On the other hand, it follows from (\ref{ON_algebra}) that the state $\sum_{a=1}^N|aa\rangle$ scatters into itself with the amplitude 
\EQ
S=NS_1+S_2+S_3\,,
\label{ON_phase}
\EN
which is equal to $-e^{3i\phi}$ for the solutions $P2_\pm$. The requirement $S=-1$ for $N=1$ (Ising free fermion) selects $P2_-$. Hence, for this solution we know $\Delta_\eta$ as a function of $N$, through (\ref{phase}), and as a function of $p$. Comparing the two results we obtain $N=2\cos\frac{\pi}{p}$. A slightly more general analysis \cite{paraf} exploiting the OPE (\ref{dnd}) with nondegenerate fields yields also $\Delta_s=\bar{\Delta}_s=\Delta_{1/2,0}$ for the dimensions of the spin field.

A rapid way to identify the solution $P2_+$ \cite{DL2} is to recall that adding the field $\Phi_{1,3}$ to the CFT with central charge (\ref{c_p}) induces a RG flow to the fixed point with central charge corresponding to $p-1$ \cite{Z_cth}. Since $\Phi_{1,3}=\varepsilon$ preserves $O(N)$ symmetry, the infrared line of fixed points obtained in this way corresponds to $P2_+$, and has $N=2 \cos\frac{\pi}{p+1}$. Together with $S=-e^{3i\phi}$, this relation leads to $\Delta_\eta=\Delta_{1,2}$, a result that differs from $\Delta_{2,1}$ for $P2_-$ for the interchange of the indices. This interchange is preserved by the mutual locality analysis based on the OPEs (\ref{dd}) and (\ref{dnd}), and yields $\Delta_\varepsilon=\Delta_{3,1}$ and $\Delta_s=\Delta_{0,1/2}$ for the solution $P2_+$. The results obtained in this way for the critical lines $P2_\pm$ are summarized in table~\ref{table_pure}. 

The critical lines $P2_\pm$ are characterized by $S_2=0$, i.e. by particle trajectories that do not intersect (see figure~\ref{ON_ampl}). It is known (see e.g. \cite{Cardy_book}) that the partition function of an $O(N)$-invariant ferromagnet can be mapped onto that of a loop gas,
\EQ
Z_\textrm{loops}=\sum_{\cal G} K^{n_b}N^{n_l}\,,
\label{loops}
\EN
where ${\cal G}$ are configurations of loops on the lattice, $K$ is the coupling in the spin formulation, $n_l$ is the number of loops, and $n_b$ is the number of edges occupied by the loops. The loop formulation implements on the lattice the continuation to noninteger values of $N$, and for $N\to 0$ is known to describe the statistics of self-avoiding walks \cite{DeGennes}. The loop model is exactly solvable on the honeycomb lattice \cite{Nienhuis}, on which the loops do not intersect. The solution produces two critical lines defined in the interval $N\in[-2,2]$ and coinciding at $N=2$. Their critical exponents were shown in \cite{DF} to correspond to the conformal dimensions $\Delta_s$ and $\Delta_\varepsilon$ that we deduced above for the solutions $P2_\pm$. The two critical lines of the loop model are referred to as ``dilute'' and ``dense'' with reference to the loop properties that they control. They correspond to the solutions $P2_-$ and $P2_+$, respectively. The analogy between particle trajectories and loop paths was originally observed in \cite{Zamo_SAW} for the off-critical case. 

\subsubsection{$N=2$ solution and the BKT phase}
The solutions $P3_\pm$ are defined only for $N=2$ and can also be written in the form
\EQ
\rho_1=\sin\alpha\,,\hspace{1cm}\rho_2=\cos\alpha\,,\hspace{1cm}\phi=-\frac{\pi}{2}\,,
\label{p3alpha}
\EN
in which $\alpha$ plays the role of the free parameter characteristic of these solutions. Its presence perfectly matches the fact, observed in section \ref{c1}, that $O(2)$ symmetry actually allows for a line of RG fixed points with central charge $c=1$. We saw that this fixed line corresponds to the Gaussian theory (\ref{free}), in which the energy density field $\varepsilon(x)=\cos 2b\varphi(x)$, with conformal dimension $\Delta_\varepsilon=b^2$, introduces the parameter $b^2$ which provides the coordinate along the line. The dimension of the chiral fields that create the particles was determined in (\ref{XYchiral}) as a function of $b^2$. On the other hand, the scattering phase (\ref{ON_phase}) takes the form $S=e^{-i\alpha}$ for the solution (\ref{p3alpha}), so that (\ref{phase}) yields the relation
\EQ
\alpha=\frac{\pi}{2b^2}\,
\label{phi_b}
\EN
between the free parameter of the scattering theory and that of the Gaussian model. Notice that $S$ takes the value $-1$ when $b^2=1/2$, in full agreement with the fact that at such a point the Gaussian model has the fermionic representation (\ref{Thirring}) with $g=0$ (free fermions). For generic $b^2$, the particles $a=1,2$ of the scattering theory can be identified with the two neutral fermions in (\ref{Thirring}). We also know that the $O(2)$ spin vector field has the bosonic representation (\ref{XYspin}), with dimension $\Delta_s=1/16b^2$.

The two intervals $\alpha\in[0,\pi/2]$ and $\alpha\in[\pi/2,\pi]$ correspond to solutions $P3_+$ and $P3_-$, respectively. As seen in section \ref{c1}, the BKT phase of the $XY$ model corresponds to the portion of the line of fixed points where $\varepsilon(x)$ is irrelevant, and then to solution $P3_+$. $P3_+$ and $P3_-$ meet at the point $\alpha=\pi/2$, which is also the meeting point of the solutions $P2_\pm$ (see figure~\ref{pure_space}). This is the BKT transition point, where the field $\varepsilon$ is marginal ($\Delta_\varepsilon=1$). 

\begin{table}
\centering
\begin{tabular}{c|c|c|c|c|c}
\hline 
Solution & $N$ & $c$& $\Delta_\eta$ & $\Delta_\varepsilon$ & $\Delta_{s}$  \\ 
\hline \hline
$\text{P}1_-$ & $(-\infty,\infty)$ & $\frac{N}{2}$ & $\frac{1}{2}$ & $\frac{1}{2}$ & $\frac{1}{16}$  \\ 
$\text{P}1_+$ & $(-\infty,\infty)$ & $N-1$ & $0$ & $1$ & $0$  \\ 
$\text{P}2_{-}$ & $2\cos\frac{\pi}{p}$ & $1-\frac{6}{p (p+1)}$  &  $\Delta_{2,1}$ & $\Delta_{1,3}$ & $\Delta_{\frac{1}{2}, 0}$ \\ 
$\text{P}2_{+}$ & $2\cos\frac{\pi}{p+1}$ & $1-\frac{6}{p(p+1)}$ &  $\Delta_{1,2}$ & $\Delta_{3, 1}$ & $\Delta_{0, \frac{1}{2}}$  \\
$\text{P}3_{\pm}$ & $2$ & $1$ & $\frac{1}{4b^2}$ &  $b^2$ & $\frac{1}{16b^2}$  \\ 
\hline 
\end{tabular} 
\caption{Central charge $c$ and conformal dimensions $\Delta$ for the solutions of table~\ref{ON_pure_solutions}. The dimensions $\Delta_{\mu,\nu}$ are given by (\ref{deltamunu}). See the text for other specifications.
}
\label{table_pure}
\end{table}

\subsubsection{Free solutions and zero temperature criticality for $N>2$}
The solutions $P1_+$ and $P1_-$ are defined for any $N$. They are purely transmissive and correspond to free bosons ($S_2=1$) and free fermions ($S_2=-1$), respectively. $P1_-$ corresponds to $N$ neutral fermions, for a total central charge $c=N/2$ (a single neutral fermion (Ising) has $c=1/2$). For $N=2$ this gives back the $c=1$ theory (\ref{free}) with $b^2=1/2$, or (\ref{Thirring}) with $g=0$; as one can see in figure~(\ref{pure_space}), this is the contact point between $P1_-$ and $P3_-$. The conformal dimension $\Delta_s=1/16$ that we give in table~\ref{table_pure} for $P1_-$ is that of the spin fields $s_1,\ldots,s_N$ of the $N$ decoupled Ising copies. Notice that, at the meeting point $b^2=1/2$, $P3_-$ has instead $\Delta_s=1/8$. The reason is that the $XY$ spin field (\ref{XYspin}) actually corresponds to $s_1s_2$ \cite{paraf}.

The solution $P1_+$ describes two different cases. On one hand, it corresponds to $N$ free bosons, i.e. the theory with action $\sum_{j=1}^N\int d^2x(\nabla\varphi_j)^2$, with $\Delta_\varepsilon=\Delta_{\varphi^2_j}=0$, and $c=N$. On the other hand, and more interestingly, for $N=2$ it also coincides with the limit $b^2\to\infty$ of $P3_+$, which has $c=1$. This is possible because, as we already observed, scattering on the line mixes statistics and interaction, so that the two fermions of the theory (\ref{Thirring}) can appear for $b^2\to\infty$ as two free bosons ($S_2=1$). This subtle role of interaction continues for $N>2$, where the $O(N)$ model is known to possess a zero temperature critical point and scaling properties described by the nonlinear sigma model (see e.g. \cite{Cardy_book})
\EQ
{\cal A}_\textrm{SM}=\frac{1}{T} \sum_{j=1}^N\int d^2x(\nabla\varphi_j)^2\,,\hspace{1cm}\sum_{j=1}^N\varphi_j^2=1\,,
\label{sm}
\EN
in which interaction is introduced by the constraint on the length of the vector $(\varphi_1, \ldots,\varphi_N)$. This theory is "asymptotically free", meaning that for $T\to 0$ the interaction amomg the bosons vanishes ($\Delta_s=\Delta_{\varphi_j}=0$), while the energy density field is marginally relevant ($\Delta_\varepsilon=1$, implying $\nu=\infty$ and exponentially diverging correlation length). The constraint in (\ref{sm}) reduces the central charge by one unit, to $c=N-1$. These results for $c$ and $\Delta_s$ agree with those for $N=2$, $b^2\to\infty$. In order to idenfity $\Delta_\varepsilon=1$, we have to observe that for $b^2>1$ at $N=2$ the field $\cos 2b\varphi$ is irrelevant, and that the most relevant $O(2)$-invariant field is the marginal one that generates the line of critical points. The sigma model interpretation of the solution $P1_+$ is the one that we report in table~\ref{table_pure}, together with the data discussed for the other solutions. 

\begin{figure}
\begin{center}
\includegraphics[width=10cm]{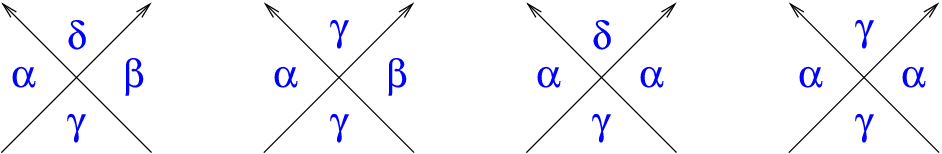}
\caption{Scattering amplitudes $S_0$, $S_1$, $S_2$ and $S_3$ for the Potts model. Time runs upwards, and different letters denote different colors.}
\label{potts_ampl}
\end{center} 
\end{figure}

\subsection{$q$-state Potts model}
\label{Potts_pure}
\subsubsection{Fixed point equations}
We have seen that the $q$-state Potts model is defined by the lattice Hamiltonian (\ref{potts}) and is characterized by invariance under permutations of the $q$ colors, corresponding to the symmetry group $\mathbb{S}_q$. The first step for the implementation of the scattering theory at criticality is again that of introducing a particle basis that carries a representation of the symmetry. For the symmety $\mathbb{S}_q$ this is achieved considering particles that we denote $A_{\alpha\beta}$, with $\alpha,\beta=1,2,\dots,q$, and $\alpha\neq\beta$. In the case of the Potts ferromagnet below critical temperature, such a particle basis corresponds to the kinks that interpolate between pairs of the $q$  degenerate vacua \cite{CZ}. As we are going to see, this same basis allows to represent the symmetry also at criticality \cite{paraf,random} (where there are no kinks due to the coalescence of the vacua) and beyond the ferromagnetic case \cite{DT1}. 

\begin{figure}
\begin{center}
\includegraphics[width=10cm]{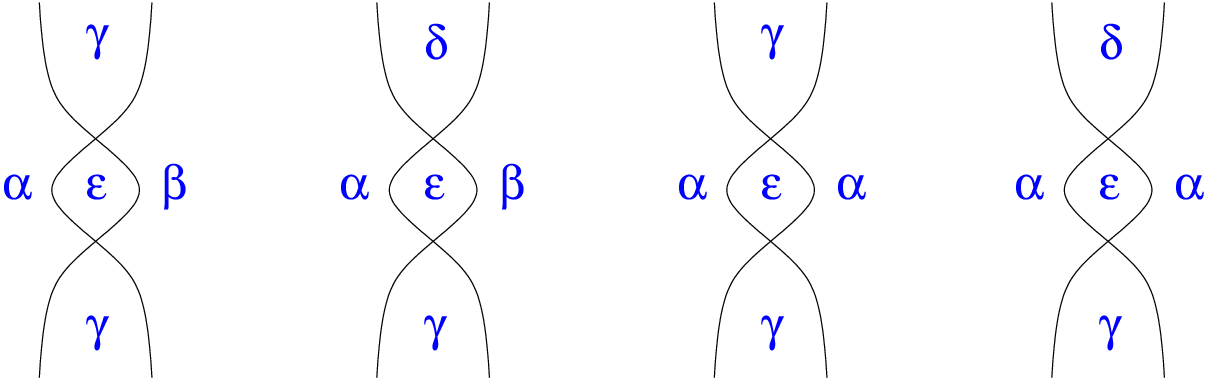}
\caption{Color configurations associated to the unitarity equations (\ref{unit1})-(\ref{unit4}), in that order. The amplitude for the lower scattering multiplies the complex conjugate of the amplitude for the upper scattering, and sum over the color $\varepsilon$ is taken.}
\label{uni_diagrams}
\end{center} 
\end{figure}

We generally think of the trajectory of a particle $A_{\alpha\beta}$ as a line separating a region of the two-dimensional space-time characterized by the color $\alpha$ from a region characterized by the color $\beta$. Permutational symmetry then allows for the four inequivalent amplitudes $S_0$, $S_1$, $S_2$ and $S_3$ shown in figure~\ref{potts_ampl}. Once the present way of labeling the particles is taken into account, the crossing relations (\ref{massless_crossing}) translate into
\bea
S_0=S_0^* & \equiv & \rho_0\,,\\
S_1=S_2^* & \equiv & \rho e^{i\phi}\,,\\
S_3=S_3^* & \equiv & \rho_3\,,
\eea
where we introduced parameterizations in terms of $\rho\geq 0$, and $\rho_0$, $\rho_3$ and $\phi$ real. In this way, the unitarity equations (\ref{massless_unitarity}) take the form \cite{paraf} (see also figure~\ref{uni_diagrams}) 
\bea
&&(q-3)\rho_0^2+\rho^2=1\,,\label{unit1}\\
&&(q-4)\rho_0^2+2\rho_0\rho\cos\phi=0\,,\label{unit2}\\
&&(q-2)\rho^2+\rho_3^2=1\,,\label{unit3}\\
&&(q-3)\rho^2+2\rho\rho_3\cos\phi=0\,.\label{unit4}
\eea
The solutions of these equations yield the Potts RG fixed points \cite{paraf,DT1}. They are listed in table~\ref{potts_solutions} and will be discussed in the next subsections. 

Notice that equations (\ref{unit1})-(\ref{unit4}) allow for a continuation of the model to noninteger values of $q$. The possibility of this  analytic continuation is well known on the lattice, where the Potts partition function allows for the expansion \cite{FK}
\EQ
Z\propto\sum_{\cal G} p^{N_b}(1-p)^{\bar{N}_b}q^{N_c}\,,
\label{FK}
\EN
where ${\cal G}$ is a graph made of bonds placed on the edges of the lattice, $N_b$ the number of bonds in ${\cal G}$, $\bar{N}_b$ the number of edges without a bond, and $N_c$ the number of clusters in ${\cal G}$, a cluster being a set of connected bonds (but also an isolated site); $p=1-e^{-J/T}$ gives the relation with the coupling $J$ of the spin representation. The Fortuin-Kasteleyn expansion is important, in particular, because shows that the percolation problem \cite{SA} can be studied as the limit $q\to 1$ of the Potts model. Indeed, in such a limit the weight $p^{N_b}(1-p)^{\bar{N}_b}$ of a bond configuration corresponds to random bond occupation with probability $p$. In the following we will discuss the Potts model implying the continuation to $q$ noninteger. While for $q$ integer only the scattering amplitudes of figure~\ref{potts_ampl} involving a number of colors not larger than $q$ play a role, all of them participate to the analytic continuation to $q$ noninteger (see \cite{DV_4point} for a detailed discussion). 

\begin{table}
\begin{center}
\begin{tabular}{c|c||c|c|c|c}
\hline
Solution & $q$ & $\rho_0$ & $\rho$ & $2\cos\phi$ & $\rho_3$ \\
\hline\hline
I & $3$ &$0$, $2\cos\phi$ & $1$ & $\in[-2,2]$ & $0$ \\ 
& & & & & \\
II$_\pm$ & $[-1,3]$ & $0$ & $1$ & $\pm\sqrt{3-q}$ & $\pm \sqrt{3-q}$\\
& & & & & \\
III$_\pm$ & $[0,4]$ & $\pm 1$ & $\sqrt{4-q}$ & $\pm\sqrt{4-q}$ & $\pm (3-q)$\\
& & & & & \\
IV$_\pm$ & $[\frac{1}{2}(7-\sqrt{17}),3]$ & $\pm \sqrt{\frac{q-3}{q^2-5q+5}}$ & $\sqrt{\frac{q-4}{q^2-5q+5}}$ & $\pm\sqrt{(3-q)(4-q)}$ & $\pm \sqrt{\frac{q-3}{q^2-5q+5}}$\\
& & & & & \\
V$_\pm$ & $[4,\frac{1}{2}(7+\sqrt{17})]$ & $\pm \sqrt{\frac{q-3}{q^2-5q+5}}$ & $\sqrt{\frac{q-4}{q^2-5q+5}}$ & $\mp\sqrt{(3-q)(4-q)}$ & $\pm \sqrt{\frac{q-3}{q^2-5q+5}}$\\
\hline
\end{tabular}
\caption{Solutions of Eqs.~(\ref{unit1})-(\ref{unit4}) yielding the RG fixed points with $\mathbb{S}_q$ permutational symmetry.} 
\label{potts_solutions}
\end{center}
\end{table}

\subsubsection{Ferromagnetic critical and tricritical lines}
In two dimensions the phase transition of the $q$-state Potts ferromagnet is known to be second order up to $q=4$, and first order above this value \cite{Baxter}. It follows that the ferromagnetic critical line corresponds to a solution of the fixed point equations (\ref{unit1})-(\ref{unit4}) having $q=4$ as upper endpoint, namely to one of the solutions III$_\pm$ in table~\ref{potts_solutions}. The fact that for $q=2$ (Ising) the only physical amplitude $S_3=\rho_3$ has to take the free fermion value $-1$ uniquely selects the solution III$_-$. 

The $q=4$ Potts model is a particular case of the Ashkin-Teller model\footnote{See \cite{AT1,DG} for the scattering description.}, which we already saw has central charge $c=1$. Hence, the Potts ferromagnetic critical line corresponds to the CFT subspace with $c\leq 1$ discussed in section~\ref{minimal}. The CFT-scattering correspondence \cite{paraf} proceeds along the same steps discussed in section~\ref{nil} for the $O(N)$ case. One first identifies $\Delta_\varepsilon=\Delta_{2,1}$, and then looks for the field $\eta$ that creates the particles as the most relevant  chiral field local with respect to $\varepsilon$; this yields $\Delta_\eta=\Delta_{1,3}$, a function of the parameter $p$ entering (\ref{c_p}). On the other hand, it follows in general from the amplitudes of figure~\ref{potts_ampl} that the state $\sum_{\gamma\neq\alpha} A_{\alpha\gamma}A_{\gamma\alpha}$ scatters into itself with amplitude
\EQ
S=S_3 + (q-2)S_2\,,
\label{eigenvalue}
\EN
which takes the value $S=e^{-4i\phi}$ for the solution III$_-$. This yields $\Delta_\eta$ as a function of $q$ through (\ref{phase}), and comparison with the previous result in function of $p$ provides the relation
\EQ
\sqrt{q}=2\cos\frac{\pi}{p+1}\,.
\label{q_cr}
\EN
The identification $\Delta_s=\Delta_{1/2,0}$ for the conformal dimension of the spin field for real values of $q$ can be done exploiting the OPE (\ref{dnd}) involving nondegenerate fields \cite{paraf}. These identifications (reported in table~\ref{summary}) of the central charge and conformal dimensions for the critical Potts ferromagnet match those obtained in \cite{DF} from the lattice determination of scaling dimensions \cite{Nienhuis}. 

We saw in section~\ref{minimal} that the critical points of the Potts ferromagnet for $q=2,3$ correspond to the CFT minimal models with $p=3,5$, respectively, and that for the same values of $q$ there are tricritical points that can be realized allowing for vacant sites and correspond to $p=4,6$, respectively. This pairing, for a given $q$, of a critical point at $p$ and and a tricritical point at $p+1$ is actually known to extend to $p$ generic (also noninteger) \cite{Z_cth}. It follows that, together with the critical line (\ref{q_cr}), there is a tricritical line with
\begin{equation}
\sqrt{q} = 2\cos\frac{\pi}{p}\,.
\label{q_trcr}
\end{equation}
The addition of vacancies does not alter color permutational symmetry, so that also the tricritical line must correspond to one of the scattering solutions in table~\ref{potts_solutions}. Critical and tricritical lines meet \cite{NBRS} at the endpoint $q=4$ ($p=\infty$), and have conformal dimensions related by index exchange (see \cite{DT1}), namely $\Delta_\varepsilon=\Delta_{1,2}$, $\Delta_s =\Delta_{0,1/2}$ and $\Delta_\eta=\Delta_{3,1}$ on the tricritical line. Inserting this value of $\Delta_\eta$ in (\ref{eigenvalue}) and using (\ref{q_trcr}) one obtains again the result $S=e^{-4i\phi}$ corresponding to III$_-$; the sign of $\sin\phi$, however, is opposite to that for the critical line\footnote{It is worth observing that the notion of "discontinuity fixed points" is used in \cite{NBRS} for the endpoints of the flow generated by renormalization group trasformations on the lattice for $q>4$, i.e. in the first order regime. In the present paper we are only concerned with fixed points existing in the continuum limit, which have $\xi=\infty$ and correspond to points of second order phase transition. See \cite{DCq4} for the exact scattering solution of the Potts model along the first order transition for $q\to 4^+$.}.

\subsubsection{Antiferromagnets}
Antiferromagnetic interaction tends to anti-align neighboring spins, and assigns to the number of neighbors an important role. Hence, while for ferromagnets quantities like critical exponents do not depend on the lattice structure (universality), antiferromagnenets have to be analyzed case by case. The fixed point equations (\ref{unit1})-(\ref{unit4}) were obtained relying only on $\mathbb{S}_q$ symmetry, which is common to ferromagnetic and antiferromagnetic Potts models. It follows that the solutions of table~\ref{potts_solutions} must also account for antiferromagnetic criticality. Actually, this observation gives a meaning to the solutions defined in ranges of $q$ other than the interval $[0,4]$ characteristic of ferromagnetic criticality. We now summarize what is presently known about the relations between the solutions of table~\ref{potts_solutions} and Potts antiferromagnets.

Solution I is defined only for $q=3$ and contains $\phi$ as free parameter. It then describes a line of fixed points, and the simplest possibility is that it corresponds to the Gaussian line with central charge $c=1$ of section~\ref{c1}. This is confirmed by the fact that the $q=3$ Potts antiferromagnet on the square lattice is known to possess a $T=0$ Gaussian critical point \cite{Baxter,LW,BH} with $\Delta_\varepsilon=b^2=3/4$ \cite{CJS}. Since we know that $\Delta_\eta=1/4b^2$ on the Gaussian line, and (\ref{eigenvalue}) gives $S=S_2=e^{-i\phi}$ for solution I, (\ref{phase}) yields the relation
\EQ
\phi=\frac{\pi}{2b^2}\,
\EN
between the scattering and the Gaussian parameters. One expects \cite{DT1} that solution I corresponds to critical points of $q=3$ antiferromagnets on lattices that change with $b^2$. Remarkably, a family of lattices (self-dual quadrangulations with $T=0$ criticality) realizing this phenomenon has recently been identified \cite{LDJS}. 

While solution I corresponds to different lattices for $q$ fixed, an early result for Potts antiferromagnets was obtained for the simplest lattice as a function of $q$. Indeed, Baxter showed that  on the square lattice there is a second order transition for $q\in[0,4]$ \cite{Baxter_square_AF}. The critical temperature decreases with $q$, and we saw a moment ago that it is zero at $q=3$, implying that for $q>3$ the transition no longer corresponds to physical temperatures. Given its range of definition, the critical line has to corresponds to one of the scattering solutions of type III, and the fact that for $q=2$ the square lattice ferromagnet and antiferromagnet can be mapped into each other selects III$_-$. The square lattice critical line was found \cite{Saleur} (see also \cite{JS,Ikhlef,DT1}) to have central charge\footnote{This value of central charge happens to be that of $\mathbb{Z}_N$ ferromagnets \cite{ZF_paraf}.}
\EQ
c=\frac{2(N-1)}{N+2}\,,
\label{c_N}
\EN
with $N$ related to $q$ as 
\EQ
\sqrt{q}=2\cos\frac{\pi}{(N+2)}\,.
\EN
One also finds $\Delta_\varepsilon=(N-1)/N$ and $\Delta_{\sigma}=N/8(N+2)$. Using (\ref{phase}) with $S=e^{-4i\phi}$ for solution III$_-$, one obtains $\Delta_\eta=2/(N+2)$. 

\begin{table}
\begin{center}
\begin{tabular}{c|c|c|c|c|c}
\hline
$\sqrt{q}$ & line & $c$ & $\Delta_{\varepsilon}$ & $\Delta_\eta$ & $\Delta_s$ \\
\hline
$2\cos\frac{\pi}{p+1}$ & F critical & $1-\frac{6}{p(p+1)}$ & $\Delta_{2,1}$ & $\Delta_{1,3}$ & $\Delta_{\frac{1}{2},0}$ \\
& & & & &  \\
$2\cos\frac{\pi}{p}$ & F tricritical & $1-\frac{6}{p(p+1)}$ & $\Delta_{1,2}$ & $\Delta_{3,1}$ & $\Delta_{0,\frac{1}{2}}$ \\
& & & & &  \\
$2\cos\frac{\pi}{N+2}$ & AF square lattice & $\frac{2(N-1)}{N+2}$ & $\frac{N-1}{N}$ & $\frac{2}{N+2}$ & $\frac{N}{8(N+2)}$ \\
\hline
\end{tabular}
\caption{Realizations of the scattering solution III$_-$ as Potts ferromagnetic (F) and antiferromagnetic (AF) critical lines. $c$ is the central charge, and the conformal dimensions $\Delta_{\mu,\nu}$ are specified by (\ref{deltamunu}).} 
\label{summary}
\end{center}
\end{table}

It follows from the latter identification that solution III$_-$ with $\sin\phi>0$ describes both the tricritical ferromagnetic line and the square lattice antiferromagnetic line, a circumstance made possible by the fact that the relation (\ref{phase}) allows for different values of $\Delta_\eta$ in correspondence of the same amplitude $S$. We then see how, remarkably, the solutions of table~\ref{potts_solutions}, while providing a compact classification of critical lines in Potts models, still allow for a variety of exponents able to account for the diversity of antiferromagnetic critical behaviors. 

The fact that for long time the known Potts critical lines \cite{Baxter_square_AF} where all defined for $q\in[0,4]$ (our solution III$_-$, see table~\ref{summary} for a summary) originated the expectation that criticality in (pure) Potts models could not be achieved for $q>4$. In this respect, our solution V is particularly interesting, since it shows that the maximal value for which criticality can occur is actually \cite{DT1}
\EQ
q_\textrm{max}=(7+\sqrt{17})/2=5.56..\,,
\label{qmax}
\EN
and leaves room for critical behavior in a $q=5$ antiferromagnet. The problem is to find a lattice on which such a critical point is realized. Actually, numerical evidence in favor of a second order transition in the five-state Potts antiferromagnet on the bisected hexagonal lattice was given in \cite{DHJSS}, but a more recent study concluded in favor of an extremely weak first order transition \cite{Salas_q5}. In this latter case the search has to start over, possibly from the families of lattices considered in \cite{HCDJKSSS}, whose numerical analysis left room for criticality at $q=5$. 

We finish the survey of the solutions of table~\ref{potts_solutions} observing that IV appears as a counterpart of V in a lower range of $q$, while II allows a conformal identification \cite{DL_tobe} similar to that of the critical lines of section~\ref{nil}.

\section{Criticality in disordered systems}
\subsection{Quenched disorder}
\label{quenched}
Statistical systems may contain "impurities" that can alter the critical properties in a very significant way. In the lattice formulation, the problem is conveniently introduced considering the case in which each site $i$ hosts, besides the usual spin variable $s_i$, a lattice gas variable $t_i$ taking the value 1 if the site is occupied (i.e. if the site contributes to the energy of that configuration of site variables), and the value 0 if the site is empty. The empty sites then play the role of the impurities. A first possibility is that the impurities are in thermal equilibrium with the  degrees of freedom of the occupied sites, in which case the sum over configurations in the partition function (\ref{pf}) is the sum over both sets of variables $\{s_i\}$ and $\{t_i\}$. This type of "annealed disorder" is not something new with respect to what we discussed in the previous sections. Indeed, we saw examples in which a suitable tuning of the density of annealed vacancies leads to tricritical behavior. The case on which we want to focus is instead that of "quenched disorder" (see e.g. \cite{Cardy_book}), in which the impurities reach thermal equilibrium on a scale of time much larger than that of the ordinary degrees of freedom. In this case the sum in the partition function is taken only over the variables $\{s_i\}$, with a "frozen" impurity configuration. This leads to a partition function $Z(\{t_i\})$. On the other hand, the impurities in such a disordered system will be randomly distributed according to some probability distribution $P(\{t_i\})$, and the free energy $F=-\ln Z$ will be averaged over the disorder,
\EQ
\overline{F}=\sum_{\{t_i\}}P(\{t_i\})F(\{t_i\})\,.
\label{Faverage}
\EN
From the theoretical point of view, a convenient a way to approach this physical situation is provided by the replica method \cite{EA}, which exploits the relation
\EQ
\overline{F}=-\overline{\ln Z}=-\lim_{n\to 0}\frac{\overline{Z^n}-1}{n}
\label{trick}
\EN
to map the problem onto that of $n\to 0$ replicas of the system coupled by the disorder average. 

It is typically possible to neglect correlations among the disorder variables $t_i$, and to treat them as independent random variables. Then the distribution
\EQ
P(\{t_i\})=\prod_i\,[p\delta_{t_i,1}+(1-p)\delta_{t_i,0}]
\label{random_vacancies}
\EN
corresponds to a concentration of impurities $1-p$. On the other hand, disorder can be associated to bonds rather than to sites. In this case, in Hamiltonians such as (\ref{vector}) or (\ref{potts}), the bond coupling $J$ is replaced by a position-dependent random bond coupling $J_{ij}$. Then, for example, the distribution
\EQ
P(\{J_{ij}\})=\prod_{\langle i,j\rangle}\,[p\,\delta_{J_{ij},J_1}+(1-p)\,\delta_{J_{ij},J_2}]\,
\label{random_bonds}
\EN
yields a mixing of randomly distributed couplings $J_1$ and $J_2$. Pure systems are recovered for $p=1$ in (\ref{random_vacancies}), and for $p=0,1$ in (\ref{random_bonds}). 

Experimental and numerical studies show that systems with quenched disorder exhibit critical properties qualitatively similar to those of pure systems. Hence, a characterization in the RG framework should again be possible. For {\it weak disorder} (e.g. $p$ only slightly smaller than 1 in (\ref{random_vacancies}) and (\ref{random_bonds})) one can start from the fixed point action ${\cal A}^*$ of the pure system and treat the effect of disorder as a perturbation, i.e. write the action
\EQ
{\cal A}={\cal A}^*+\int d^dx\,m(x)\varepsilon(x)\,,
\label{A_random}
\EN
where $\varepsilon(x)$ is the energy density field of the pure system. The difference with respect to (\ref{scaling}) is that now the coupling is a position dependent random variable $m(x)$ obeying a distribution $P(\{m(x)\}$. Then, within the replica framework, one has to consider
\EQ
\overline{Z^n}=\sum_{\{m(x)\}}P(\{m(x)\} \sum_{\textrm{field configurations}}e^{-\sum_a[{\cal A}_a^*+\int d^dx\,m(x)\varepsilon_a(x)]}\,,
\EN
where $a=1,\ldots,n$ is the index labeling the different replicas. A cumulant expansion of $P(\{m(x)\}$ (see \cite{Cardy_book}) then leads to 
\EQ
\overline{Z^n}=\sum_{\textrm{field configurations}}e^{-\sum_a[{\cal A}_a^*+g_1\int d^dx\,\varepsilon_a(x)]+g_2\sum_{a\neq b}\int d^dx\,\varepsilon_a(x)\varepsilon_b(x)+\cdots}\,,
\label{effective}
\EN
with the couplings $g_1$ and $g_2$ related to the first two cumulants for uncorrelated disorder. Hence, we see that the leading effect of weak disorder is that of coupling different replicas through the field $\varepsilon_a\varepsilon_b$. Since the scaling dimension of the latter is twice the scaling dimension $X_\varepsilon$ of the energy density field {\it at the pure fixed point}, one concludes that weak disorder is irrelevant in the RG sense if 
\EQ
2X_\varepsilon>d\,.
\label{Harris}
\EN
This is known as Harris criterion \cite{Harris} and, since the specific heat critical exponent $\alpha$ is given by $(d-2X_\varepsilon)\nu$, it is more often quoted in the form $\alpha<0$. If (\ref{Harris}) is satisfied, weak disorder does not change the critical properties of the system. On the contrary, if $2X_\varepsilon<d$, disorder is relevant and, even in small concentration, can alter the large distance properties driving the system to a new, "random", RG fixed point with critical exponents differing from those of the pure case. The generality of (\ref{effective}) indicates that the critical exponents of this new fixed point should not depend on the type of disorder (e.g. site or bond).

Harris criterion, on the other hand, says nothing about critical points that can be produced by strong disorder, or about the effect of disorder when the transition in the pure system is first order. We will now see how the scattering framework gives general and exact access to random criticality in $d=2$.

\subsection{Disordered Potts model}
\subsubsection{Fixed point equations}
The two-dimensional random bond $q$-state Potts model, defined by the lattice Hamiltonian
\EQ
{\cal H}=-\sum_{\langle i,j\rangle}J_{ij}\delta_{s_i,s_j}\,,
\EN
with $s_i=1,2,\ldots,q$, has played a particularly relevant role in the theoretical study of quenched disorder in systems with short range interactions. We saw that the pure ferromagnet ($J_{ij}=J>0$) has a phase transition that is second order up to $q=4$ and becomes first order above this value. It follows from the value of $X_\varepsilon=2\Delta_\varepsilon$ given in table~\ref{summary} for the critical line of the pure ferromagnet that weak disorder is marginal (actually marginally irrelevant \cite{DD}) at $q=2$ and relevant for $q\in]2,4]$. Slightly above $q=2$ disorder is weakly relevant, a condition that generally allows the determination of a new perturbative fixed point (see \cite{Cardy_book}). In the present case the perturbative analysis was performed in \cite{Ludwig,DPP} and leads to a random fixed point that, while perturbatively defined for $q\to 2^+$, is expected to persist as long as the Harris criterion can be applied, i.e. up to the endpoint $q=4$ of the pure critical line. On the other hand, it is known \cite{AW,HB} that disorder softens the first order transition that the pure model exhibits for $q>4$, and that a second order transition extending to infinite $q$ has to be expected. In other words, the random critical line originating at $q=2$ should persist for all larger values of $q$, a circumstance confirmed by numerical investigations. Early numerical \cite{CFL,DW,KSSD} and experimental \cite{random_exp} studies also suggested a peculiar "superuniversality" (i.e. $q$-independence) of critical exponents along this critical line, a scenario no longer considered after that clearly $q$-dependent data for the order parameter exponent $\beta$ were obtained in \cite{CJ}. 

\begin{figure}
\begin{center}
\includegraphics[width=10cm]{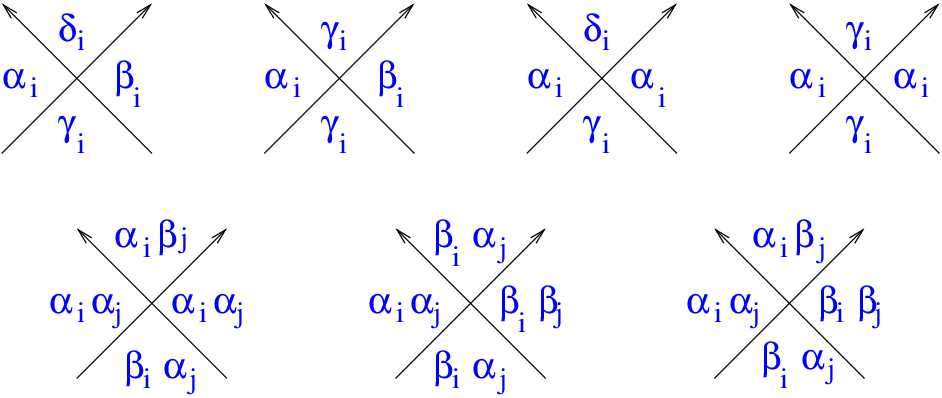}
\caption{Scattering amplitudes $S_0$, $S_1$, $S_2$, $S_3$, $S_4$, $S_5$, $S_6$ in the replicated $q$-state Potts model, in that order. Time runs upwards. Different replicas are indicated by different latin indices, and different colors for the same replica by different greek letters. 
}
\label{random_potts_ampl}
\end{center} 
\end{figure}

Progress is allowed by the fact that the scale invariant scattering framework used in section~\ref{Potts_pure} for the pure model can be naturally generalized to the random case \cite{random}. Following the general logic of the previous section, we consider $n$ replicas of the system, each one with its own excitations. For the Potts model these will be the particles $A_{\alpha_i\beta_i}$ ($\alpha,\beta=1,2,\ldots,q$; $\alpha\neq\beta$) for replica $i=1,2,\ldots,n$. Generalizing what we saw for the pure case ($n =1$), the trajectory of a particle $A_{\alpha_i\beta_i}$ now separates a region of the plane characterized by the colors $\alpha_1, \ldots,\alpha_n$ for the replicas $1, \dots,n$, respectively, from a region in which replica $i$ changes its color to $\beta_i$, while the colors of the other replicas do not change. It follows that the scattering amplitudes inequivalent under the requirement of invariance under permutations of the replicas and permutations of the colors within each replica are those shown in figure~\ref{random_potts_ampl}. For simplicity, only the replicas whose color changes in the scattering are indicated in the figure. The amplitudes $S_{k\leq 3}$ involve color change within a single replica, and are the analog of those of figure~\ref{potts_ampl} for the pure model. The last three amplitudes involve interaction between different replicas and are characteristic of the disordered case. For example, the amplitude $S_4$ corresponds to the process in which the initial state with particles $A_{\alpha_i,\beta_i}$ and $A_{\beta_i,\alpha_i}$, both in replica $i$, scatters into the final state with particles $A_{\alpha_j,\beta_j}$ and $A_{\beta_j,\alpha_j}$, both in replica $j$.

\begin{figure}
\begin{center}
\includegraphics[width=15cm]{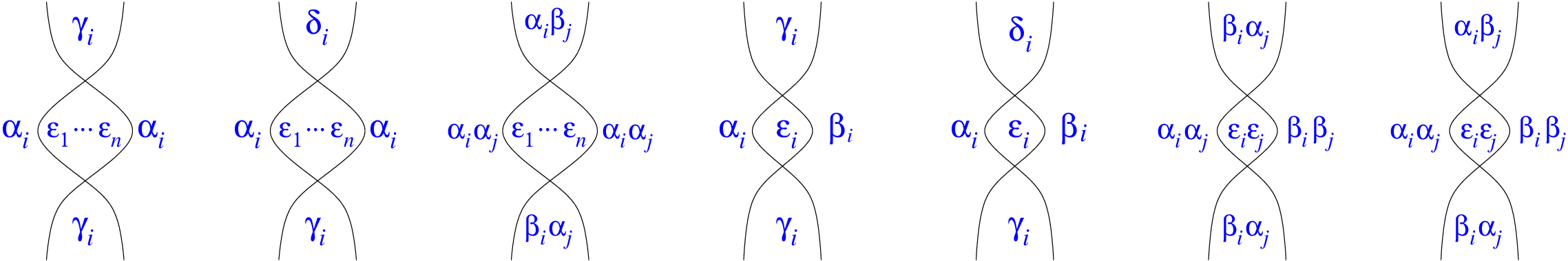}
\caption{The different color configurations that give rise to the unitarity equations (\ref{u1})--(\ref{u6}), in the same order. The amplitude for the lower scattering multiplies the complex conjugate of the amplitude for the upper scattering, and colors in the closed region are summed over. Those of the $n$ replicas that are not indicated keep the same color in the four external regions and, for the last four diagrams, also in the internal region.
}
\label{random_unitarity}
\end{center} 
\end{figure}

As usual, crossing symmetry (\ref{massless_crossing}) provides a relation between amplitudes under  the exchange of time and space directions, and in this case takes the form
\bea
S_0 = S_0^* & \equiv & \rho_0\,,
\label{cr1}\\
S_1 = S_2^*& \equiv &\rho\,e^{i\phi}\,,
\label{c2}\\
S_3 = S_3^*& \equiv &\rho_3\,,
\label{c3}\\
S_4 = S_5^*& \equiv &\rho_4\,e^{i\theta}\,,
\label{c4}\\
S_6 = S_6^*& \equiv &\rho_6\,,
\label{c5}
\eea
where we introduced the parameterizations in terms of $\rho_0, \rho_3,\rho_6, \phi,\theta\in\mathbb{R}$ and $\rho, \rho_4\geq 0$. The unitarity equations (\ref{massless_unitarity}) are now associated to the color configurations shown in figure~\ref{random_unitarity}, which generalize to the present replicated case those of figure~\ref{uni_diagrams} for the pure model. They read~\cite{random} 
\bea
&& \rho_3^2+(q-2)\rho^2+(n-1)(q-1)\rho_4^2 = 1\,,
\label{u1}\\
&& 2\rho\rho_3\cos\phi+(q-3)\rho^2+(n-1)(q-1)\rho_4^2 = 0\,,
\label{u2}\\
&& 2\rho_3\rho_4\cos\theta+2(q-2)\rho\rho_4\cos(\phi+\theta)+(n-2)(q-1)\rho_4^2 = 0\,
\label{u3}\\
&& \rho^2+(q-3)\rho_0^2 = 1\,,
\label{u4a}\\
&& 2\rho_0\rho\cos\phi+(q-4)\rho_0^2 = 0\,,
\label{u4b}\\
&& \rho_4^2+\rho_6^2=1 \,,
\label{u5}\\
&& \rho_4\rho_6\cos\theta=0 \,.
\label{u6}
\eea
Notice that, besides $q$, now also the number $n$ of replicas enters the equations as a parameter that can take real values, so that the limit $n\to 0$ required by (\ref{trick}) can be taken straighforwardly.

As expected, when $n$ is set to 1 and the equations still involving $\rho_4$ or $\rho_6$ are ignored, the equations (\ref{u1})--(\ref{u6}) reduce to those (\ref{unit1})-(\ref{unit4}) of the pure model. Another relevant observation is that $\rho_4=0$ leads to $n$ noninteracting replicas, since it implies $S_4=S_5=0$ and, as a consequence of (\ref{u5}), $S_6=\pm 1$. Since we saw that the replicas are coupled by the disorder, this means that $\rho_4=0$ corresponds to absence of disorder. 

The solutions of the equations (\ref{u1})--({\ref{u6}), which for $n=0$ and $\rho_4\neq 0$ give the RG fixed points of the disorderd Potts model, have been listed in \cite{DT2}. Since the equations are exact, they yield not only the fixed points produced by weak disorder, but also those beyond the predictivity of Harris criterion. In addition, since the equations generally implement $\mathbb{S}_q$ permutational symmetry, they yield the random fixed points associated to all the solutions of the pure case (table~\ref{potts_solutions}), including the antiferromagnetic ones. In the following, we will discuss only the case in which absence of disorder corresponds to the pure ferromagnet.

\subsubsection{Softening and superuniversality}
We know that the critical line of the pure ferromagnet extends up to $q=4$ and corresponds to the solution III$_-$ of table~\ref{potts_solutions}. We saw in the previous subsection that a combination of perturbative, rigorous and numerical results leads to the expectation of a line of random fixed points extending from $q=2$ (where it corresponds to pure Ising, i.e. $\rho_3=-1$ and $\rho_4=0$) to infinite $q$. The equations (\ref{u1})-(\ref{u6}) with $n=0$ indeed possess such a solution, which reads \cite{random}
\EQ
\rho _0=\cos\theta=0,\hspace{1cm} \rho =1,\hspace{1cm} \rho _3=2 \cos\phi =-\frac{2}{q}, \hspace{1cm}\rho _4=\frac{q-2}{q} \sqrt{\frac{q+1}{q-1}}\,,
\label{ir}
\EN
and provides the first analytic verification of the above expectation (see figure~\ref{soft_pd}).

\begin{figure}
\begin{center}
\includegraphics[width=8cm]{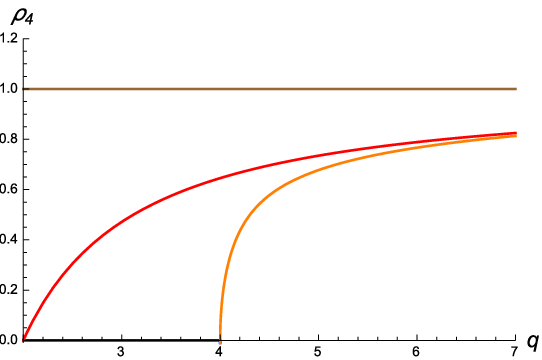}
\caption{The parameter $\rho_4$ for the solutions of the equations (\ref{u1})-(\ref{u6}) with $n=0$ discussed in the text and corresponding to lines of RG fixed points in the disordered $q$-state Potts model. Absence of disorder amounts to $\rho_4=0$, and the critical line of the pure ferromagnet is shown. The critical lines (\ref{ir}) and (\ref{uv}) with varying $\rho_4$ originate at $q=2$ and $q=4$, respectively.  Nishimori-like and zero temperature critical points have $\rho_4=1$.  
}
\label{soft_pd}
\end{center} 
\end{figure}

From the RG point of view, the critical line (\ref{ir}) is the large distance limit of a flow with infinite correlation length, which for $q\in(2,4]$ originates from the critical line of the pure ferromagnet. However, since (\ref{ir}) extends to $q=\infty$, it should be possible to identify a starting point of the flow also for $q>4$. Since in this range the pure ferromagnet has a first order transition (finite correlation length), a line of unstable RG fixed points for $q>4$ should exist in the disordered system. Given the generality of the fixed point equations (\ref{u1})-(\ref{u6}), they should then admit a solution starting at $q=4$ and extending until $q=\infty$. This solution indeed exists and reads \cite{DL_softening}
\bea
&& \rho _0 =-\frac{2}{q^2-4 q+2}\,, \hspace{1cm}
\rho =\frac{\sqrt{(q-4) (q^3-4 q^2+4 q-4)}}{q^2-4 q+2}\,,\nonumber\\
&& \rho _3 =\frac{2(q-3)}{q^2-4 q+2}\,, \hspace{1cm}
\rho _4 =\frac{(q-2)\sqrt{(q-4) (q-3) (q-1) q}}{(q-1)(q^2-4 q+2)}\,, \nonumber\\
&& 2 \cos\phi =\frac{2 (q-4)}{\sqrt{(q-4) (q^3-4 q^2+4 q-4)}}\,,\hspace{1cm}\cos\theta=0\,.
\label{uv}
\eea
Notice that at $q=4$ this solution coincides with that for the pure ferromagnet, i.e. III$_-$ of table~\ref{potts_solutions} (see figure~\ref{soft_pd}). The existence of the solution (\ref{uv}) suggests that for $q>4$ the softening of the first order transition of the pure ferromagnet into the second order transition corresponding to (\ref{ir}) sets in only above a $q$-dependent disorder threshold. Below this threshold the transition would remain first order (in the sense of a finite correlation length) but, in order to comply with the rigorous result of \cite{AW}, without the usual discontinuity in the energy density (latent heat). In this respect, it must be remarked that a first order transition without latent heat was found in mean field theory for $q>4$ in a Potts model with quenched disorder \cite{GKS}. 

\begin{figure}
\begin{center}
\includegraphics[width=10cm]{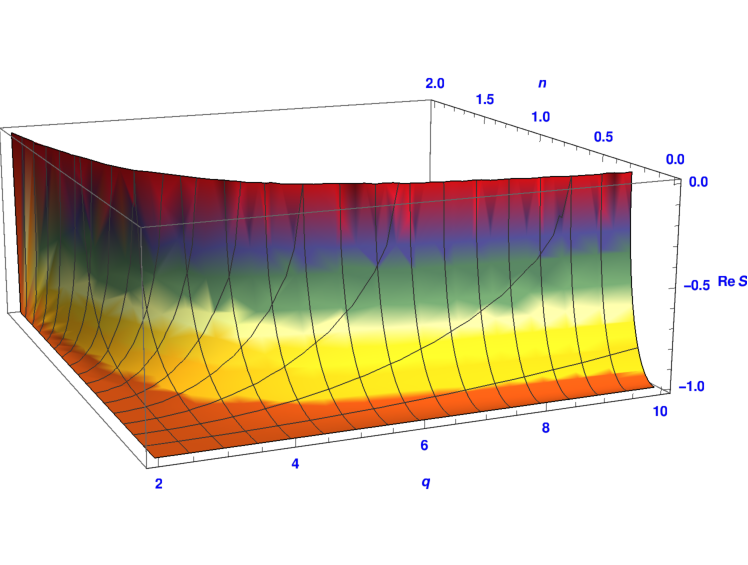}
\caption{The scattering phase (\ref{S_random_potts}) becomes $q$-independent in the limit $n\to 0$ yielding the critical line (\ref{ir}).
}
\label{S(q,n)}
\end{center} 
\end{figure}

The critical lines (\ref{ir}) and (\ref{uv}) have $\cos\theta=0$, a property that has a remarkable implication. Indeed, the combination $\sum_{i,\gamma_i}A_{\alpha_i\gamma_i}A_{\gamma_i\alpha_i}$ of two-particle states is invariant under color permutations ($\mathbb{S}_q$-invariance) and scatters into itself with the phase
\EQ
S= S_3+(q-2)S_2+(n-1)(q-1)S_4\,,
\label{S_random_potts}
\EN
which generalizes (\ref{eigenvalue}) to the replicated case. It is easily checked that for $\cos\theta=0$ equation (\ref{u3}) implies $\textrm{Im}\,S=0$ (namely $S=\pm 1$) in the case $n=0$ of interest for quenched disorder (see figure~\ref{S(q,n)}). This means that, while in general the solutions (\ref{ir}) and (\ref{uv}) depend on $q$, their $\mathbb{S}_q$-invariant sector does not, an independence on the symmetry parameter that does not occur in pure systems. It makes possible that the scaling dimensions of fields belonging to the $\mathbb{S}_q$-invariant sector, in particular the energy density $\varepsilon(x)$, stay constant along the random critical line. This allows the exponent $\nu=1/(2-X_\varepsilon)$ to preserve along the critical line (\ref{ir}) the pure Ising value $1$ that it takes at $q=2$, a circumstance that sheds light on the persisting indications \cite{CFL,DW,KSSD,random_exp,CJ,CB2,OY,JP,Jacobsen_multiscaling,AdAI} that $\nu$ does not exhibit appreciable deviations from its Ising value up to $q=\infty$. We observe that the value $\nu=1$ saturates the lower bound $\nu=2/d$ obtained in \cite{CCFS} for disordered systems in $d$ dimensions. On the other hand, the order parameter critical exponent $\beta$ depends on the scaling dimension $X_s$ of the spin field, which is not $\mathbb{S}_q$-invariant; hence $\beta$ has to be $q$-dependent, again in agreement with the numerical determinations of \cite{CJ} and subsequent studies. Hence, the exact fixed point equations (\ref{u1})-(\ref{u6}) allow for a subtle mechanism of superuniversality of some critical exponents \cite{random}. We will see that this mechanism is not limited to the Potts model and appears to be characteristic of quenched disorder. 

\begin{figure}
\begin{center}
\includegraphics[width=6cm]{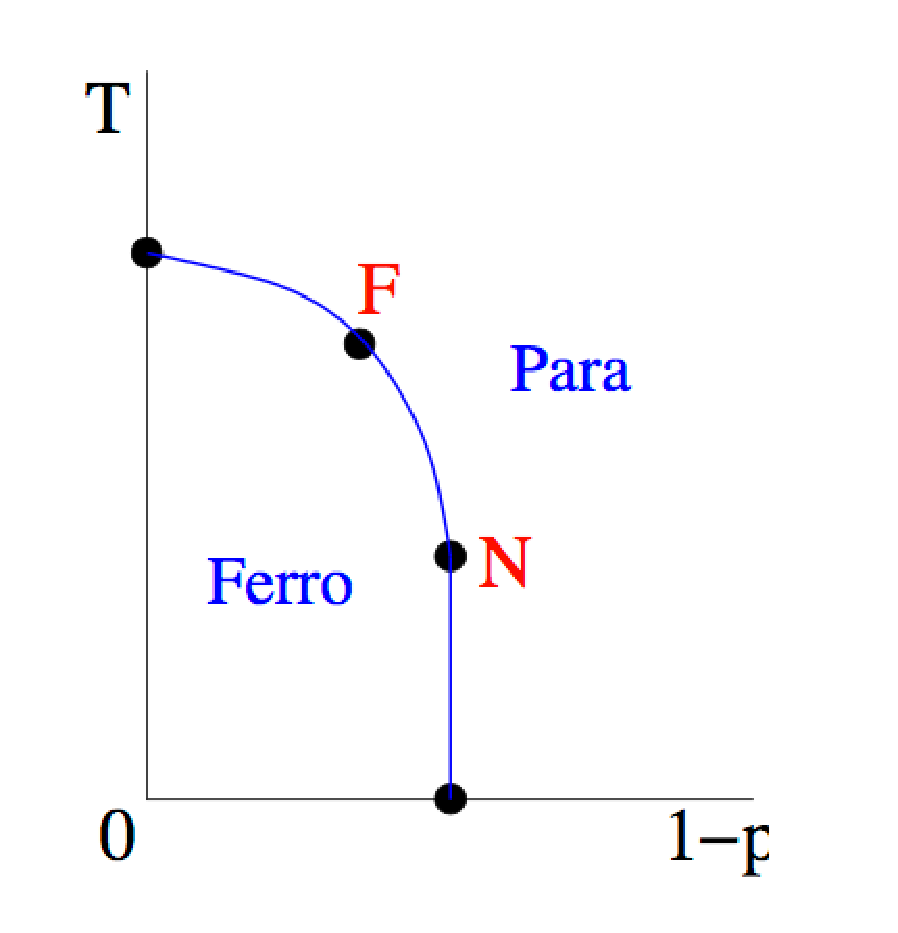}
\caption{Qualitative phase diagram and expected RG fixed points (dots) in the two-dimensional three-state Potts model for the random bond distribution (\ref{random_bonds}) with $J_1=-J_2>0$. $F$ is the fixed point (\ref{ir}) to which the system is driven by weak disorder; $N$ denotes the Nishimori-like fixed point (\ref{nishimori}). 
}
\label{random_q3}
\end{center} 
\end{figure}

The critical line (\ref{ir}) rules the large distance properties of the system obtained adding weak disorder to the pure ferromagnet. On the other hand RG fixed points corresponding to stronger disorder are also expected and numerically observed (see \cite{PHP} and references therein for $q=2$). For the disorder distribution (\ref{random_bonds}) with $J_1=-J_2>0$, in which disorder is associated to antiferromagnetic bonds, one such a strong disorder fixed point can be referred to as Nishimori-like fixed point, since it generalizes the Nishimori fixed point that at $q=2$ can be argued from a lattice gauge symmetry \cite{Nishimori}; for $q=3$ it has been studied numerically in \cite{SGH,JP2} (see figure~\ref{random_q3}). The solution of equations (\ref{u1})-(\ref{u6}) corresponding to Nishimori-like fixed points was identified in \cite{DT2} with 
\EQ
\rho _0=0,\hspace{1cm} \rho=\rho_4=1,\hspace{1cm}\rho_3=2\cos\phi =-\sqrt{2}, \hspace{1cm} 
2\cos\theta = -\frac{\sqrt{2}(q^2-2)}{\left(q^2-2 q+2\right)}\,,
\label{nishimori}
\EN
and belongs to the class of solutions with $\rho_4=1$, the only one for which disorder never becomes weak as $q$ varies. This class also contains the solution 
\EQ
\rho _0=0,\hspace{1cm}\rho=\rho_4=1,\hspace{1cm} \rho_3=2\cos\phi=2\cos\theta =-\sqrt{2}\,,
\label{perc}
\EN
characterized by complete $q$ independence, a feature expected within the space of solutions. Indeed, if we consider a bond diluted ferromagnet, namely the disorder distribution (\ref{random_bonds}) with $J_1=J>0$ and $J_2=0$, only islands made of spins connected by bonds $J_{ij}=J$ contribute to the energy. At zero temperature these islands are uniformly and independently colored, and give a zero order parameter unless there is an infinite cluster. Hence, a percolation transition is expected for any $q$, and is accounted for by (\ref{perc}).

\subsection{Disordered $O(N)$ model}
\subsubsection{Fixed point equations}
The vector model with random bonds is defined on the lattice by the Hamiltonian
\EQ
{\cal H}=-\sum_{\langle i,j\rangle}J_{ij}\,{\bf s}_i\cdot{\bf s}_j\,,
\label{lattice}
\EN
where ${\bf s}_j$ is a $N$-component unit vector located at site $j$. The scattering formulation at critical points in two dimensions again exploits the replica method, and generalizes that of the pure model taking into account that vector multiplets of particles now exist in each replica. We then denote the particles as ${a_i}$, with $a=1,2,\ldots N$ labeling the components of the vector multiplet, and $i=1,2,\ldots,n$ labeling the replicas. As in the pure case, the scattering yields annihilation, transmission and reflection processes. These, however, can now occur in a single replica (amplitudes $S_1,S_2,S_3$ of figure~\ref{random_vector_ampl}) or in different replicas (amplitudes $S_4, S_5, S_6$). Given these amplitudes, crossing symmetry (\ref{massless_crossing}) yields the relations and parameterizations
\bea
S_1=S_3^{*} & \equiv &  \rho_{1}\,e^{i\phi}, \\
S_2 = S_2^* & \equiv & \rho_2,\\
S_4 = S_6^* & \equiv &  \rho_4\, e^{i\theta}, \\
S_5 = S_5^*& \equiv & \rho_5\,.
\eea 
On the other hand, the unitarity equations (\ref{massless_unitarity}) take the form \cite{DL1}
\bea
&& \rho_1^2+\rho_2^2=1\,,  \label{uu1}\\
&& \rho_1 \rho_2 \cos\phi=0\,, \label{uu2}  \\
&& N \rho_1^2 + N(n-1)\rho_4^2 + 2\rho_1\rho_2 \cos\phi +2\rho_1^2\cos2\phi=0\,, \label{uu3} \\
&& \rho_4^2 + \rho_5^{2}=1\,, \label{uu4}\\
&& \rho_4 \rho_5\cos\theta=0\,, \label{uu5}\\
&& 2 N \rho_1 \rho_4 \cos(\phi-\theta) + N(n-2)\rho_4^2+2\rho_2\rho_4\cos\theta + 2\rho_1\rho_4\cos(\phi+\theta)=0\, \label{uu6}. 
\eea
As we by now expect, $N$ and $n$ enter the equations as parameters that do not need to take integer values; in particular, the limit $n\to 0$ required by the replica approach to quenched disorder can be taken without difficulties. Notice also that the case $\rho_4=0$ corresponds to decoupled replicas ($S_5=\pm 1$), and then to absence of disorder. 

\begin{figure}
\begin{center}
\includegraphics[width=16cm]{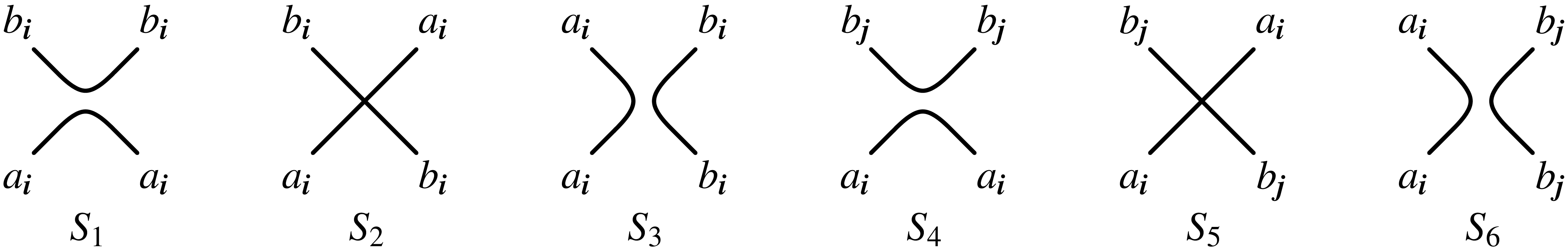}
\caption{Scattering amplitudes for the replicated $O(N)$ vector theory. Time runs upwards, and the indices $i$ and $j$ label different replicas. 
}
\label{random_vector_ampl}
\end{center} 
\end{figure}

\subsubsection{Critical lines}
The solutions of equations (\ref{uu1})-(\ref{uu6}) were given in \cite{DL2}, and for $n=0$, $\rho_4\neq 0$ provide the RG fixed points of the disordered $O(N)$ model. We now discuss the solutions corresponding to critical lines arising when disorder is added to the pure ferromagnet. We know from table~\ref{table_pure} that the latter corresponds to solution P2$_-$ for $N<2$, and to solution P1$_+$ for $N>2$. We also know from the same table that $2X_\varepsilon=4\Delta_\varepsilon<2$, so that weak disorder is relevant, only for $N<1$. In addition, disorder becomes weakly relevant as $N\to 1^-$, so that a random fixed point can be found perturbatively in this limit; this perturbative analysis was performed in \cite{Shimada}. Hence, below $N=1$, weak disorder drives the system to a new critical line that merges the solution P2$_-$ of the pure system at the Ising point. The equations (\ref{uu1})-(\ref{uu6}) yield this new critical line, which reads \cite{DL1}
\EQ
\rho_1=1,\quad \rho_2=\cos\theta=0,\quad \cos\phi=-\frac{1}{N+1},\quad \rho_4=\frac{1-N}{1+N}\sqrt{\frac{N+2}{N}}\,,
\label{varying}
\EN
and extends from $N=1$, where $\rho_4=0$, down to
\EQ
N_*=\sqrt{2}-1= 0.414..\,,
\label{nstar}
\EN
where $\rho_4$ reaches its maximal value 1 (see (\ref{uu4}) and figure~\ref{ON_modulus}). The presence of a lower endpoint $N_*$ for this critical line had been argued perturbatively in \cite{Shimada}, where the estimate $N_*\approx 0.26$ had been obtained in a two-loop approximation. The subsequent  numerical estimate $N_*\approx 0.5$ obtained in \cite{SJK} is not far from the exact result (\ref{nstar}).

\begin{figure}
\begin{center}
\includegraphics[width=8cm]{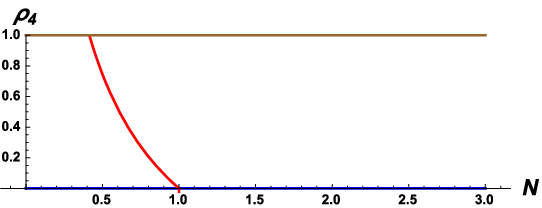}
\caption{The parameter $\rho_4$ for the solutions of the equations (\ref{uu1})-(\ref{uu6}) with $n=0$ discussed in the text. Absence of disorder ($\rho_4=0$) corresponds to the critical $O(N)$-invariant pure ferromagnet. The critical line (\ref{varying}) with varying $\rho_4$ is expected from Harris criterion below $N=1$, and is found to terminate at $N_*=\sqrt{2}-1$. Nishimori-like and zero temperature critical points have $\rho_4=1$.  
}
\label{ON_modulus}
\end{center} 
\end{figure}

Notice that the superposition $\sum_{a,i}a_ia_i$ of two-particle states, which is invariant under $O(N)$ transformations, scatters into itself with amplitude
\EQ
S=NS_1+S_2+S_3+(n-1)NS_4\,,
\label{singlet}
\EN
which generalizes (\ref{ON_phase}) to the replicated case. It is easily checked that, when evaluated for the $n=0$ solution (\ref{varying}), $S$ becomes $N$-independent. Hence, similarly to what we saw in the disordered Potts model, also in the $O(N)$ case we have a critical line along which a phenomenon of superuniversality appears. In particular, the correlation length critical exponent $\nu$ is expected to be $N$-independent along this critical line, and to preserve the value 1 that it has at $N=1$. On the other hand, the order parameter field is not $O(N)$-invariant and does not belong to the superuniversal sector. Its scaling dimension $X_s$ has to vary along the critical line, and this is consistent with the measurement of \cite{SJK} at $N=0.55$. 

At $N=N_*$ the critical line (\ref{varying}) merges the solution 
\EQ
\rho_1=\rho_4=1, \quad\rho_2=0,\quad\cos\phi=-\frac{1}{\sqrt{2}},\quad \cos\theta=-\frac{N^2+2N-1}{\sqrt{2}(N^2+1)} \,,
\label{strong}
\EN
which is defined for any $N$ and is always strongly disordered ($\rho_4=1$, see figure~\ref{ON_modulus}). It is then expected that at $N=1$ this solution corresponds to the Nishimori point of the Ising model. A line of Nishimori-like RG fixed points with these properties was indeed observed numerically in \cite{SJK}. Additional features of the critical lines (\ref{varying}) and (\ref{strong}), as well as of zero temperature criticality, are discussed in \cite{DL2}.

\section{Conclusion}
In this article we reviewed progress that has been recently achieved for two-dimensional systems at criticality implementing conformal invariance within the particle description of the underlying field theory. The infinite-dimensional character of conformal symmetry in $d=2$ induces essential simplifications in the scattering formalism, which then yields exact equations whose solutions provide a classification of RG fixed points with a given internal symmetry. We illustrated the method for two main models of the theory of critical phenomena, namely the $O(N)$ vector model and the $q$-state Potts model, for which critical lines are obtained as the symmetry parameters $N$ and $q$ are varied (see \cite{DL_XYising,DDL} for the study of other symmetries). In the case of pure systems, for which many exact results are already known, the formalism allows to obtain the different critical lines with the given symmetry from a single set of equations, and to gain a global view of their location in the space of parameters. In addition, previously unknown critical points can appear; they are most often related to antiferromagnets, for which a first global viewpoint becomes available. 

Remarkably, the same type of analysis can be extended to the systems with quenched disorder, for which the relevance of conformal invariance had seemed too hard to substantiate. The scattering approach finally gives exact access to random criticality and, again, provides a global view in which the space of solutions of the fixed point equations contains the critical lines that had been studied perturbatively in special limits of weak disorder, as well as the strong disorder critical lines (Nishimori-like or zero-temperature) that had been completely out of reach of analytical methods. Phenomena such as the softening of first order transitions by disorder are also observed analytically for the first time. 

One striking result emerging from this exact exploration of random criticality is the presence of lines of fixed points along which {\it some} critical exponents can stay constant as the symmetry parameter is changed. This superuniversality mechanism sheds light on numerical observations that had gathered over the years and had remained unclear in absence of a theoretical picture. While the phenomenon has no counterpart in pure systems, it appears to be generic for disordered ones in $d=2$. At the same time, it is hard to point out a reason why it should be confined to two dimensions. Experimental \cite{ICZDF} and numerical \cite{UHJ,Parisi_etal,Janke_etal,MAI} studies show that also in $d=3$ disorder is able to soften the first order transition of a pure system into a second order one, again making room for new conformally invariant points. Additional numerical work aimed at verifying the symmetry dependence of critical exponents in $d=3$, for example for the $q$-state Potts model as a function of $q$, would be very interesting.

For the two-dimensional case, now that the relevance of conformal invariance for random criticality has been established exploiting the particle framework, it will be relevant to understand how random RG fixed points can also be described within the more traditional framework of the infinite-dimensional conformal algebra. 

We finally observe that the relevance of conformal invariance is well known also for surface critical behavior in pure systems \cite{Cardy_surface}. It will be interesting to extend the scale invariant scattering approach to systems with a boundary\footnote{Away from criticality, boundary scattering in two dimensions is known to admit exact solutions \cite{GZ} and has been applied, in particular, to percolation \cite{DV_crossing} and the wetting transition \cite{DS_wetting,localization}.} and gain access to surface criticality in presence of quenched disorder.



\end{document}